\documentclass[galaxies,article,submit,pdftex,moreauthors]{Definitions/mdpi} 

\firstpage{1} 
\pubvolume{1}
\issuenum{1}
\articlenumber{0}
\pubyear{2024}
\copyrightyear{2024}
\datereceived{ } 
\daterevised{ } 
\dateaccepted{ } 
\datepublished{ } 
\hreflink{https://doi.org/} 

\Title{Examination of the functional form of the light and mass distribution in spiral arms}

\TitleCitation{Examination of the light and mass distribution in spiral arms}

\Author{Ilia V. Chugunov $^{1}$*\orcidA{}, Alexander A. Marchuk $^{1, 2}$\orcidB{}, Sergey S. Savchenko$^{1, 2}$\orcidC{}}

\AuthorNames{Ilia V. Chugunov, Alexander A. Marchuk, Sergey S. Savchenko}

\AuthorCitation{Chugunov I.; Marchuk A.; Savchenko S.}

\address{%
$^{1}$ \quad Pulkovo Astronomical Observatory, Russian Academy of Sciences, St. Petersburg 196140, Russia\\
$^{2}$ \quad St. Petersburg State University, Universitetskii pr. 28, St. Petersburg 198504, Russia}

\corres{Correspondence: chugunov21@list.ru}

\abstract{Spiral arms are a common feature of local galaxies, but the exact form of the distribution of mass and light in them is not well known. In this work, we aim to measure this distribution as accurately as possible, focusing on individual spiral arms and using the so-called slicing method. The sample consists of 19 well-resolved, viewed face-on spiral galaxies from the S$^4$G survey. We work primarily with infrared images at 3.6 $\mu$m from the same survey, and, secondarily, with ultraviolet data from the GALEX telescope. We derive the properties of the spiral arms step by step, starting from their overall shape, then measuring their brightness profile and width variation along the arm, and then examining the fine structure of the profile across the arm, namely its skewness and S{\'e}rsic index. We construct a 2D photometric function of the spiral arm that can be used in further decomposition studies, validate it and identify the most and least important parameters. Finally, we show how our results can be used to unravel the nature of the spiral arms, supporting the evidence that NGC~4535 has a density wave in its disc.}


\begin{document}


\section{Introduction}

Spiral galaxies are the most common type of bright galaxies in the local Universe. 75\% of galaxies brighter than $M_B = -20$ are spiral, including our Milky Way~\cite{Conselice2006}. However, the nature and properties of spiral structure are still not well-known (see review~\cite{Dobbs2014} and references therein).

Spiral structure stands out aganist the disc background by higher surface brightness, bluer colours and the concentration of younger objects. In the same time, spiral structure is known to be interconnected with various fundamental properties of galaxies. Simulations suggest that spiral arms can be traced by other indicators like metallicity or dynamical properties~\cite{Breda2024, Debattista2025}. Another example is that star formation is clearly enhanced in spiral arms, but the exact form of this connection is still a matter of discussion~\cite{Yu2021, Querejeta2021,Querejeta2024}. As for our Galaxy, our location inside the Milky Way has some advantages, allowing to examine its spiral structure in detail and understand its physical nature more. As such, our current state of knowledge about spiral arms in our Galaxy is more advanced than for outer galaxies in some aspects. For example, spiral structure, especially in the solar vicinity, was mapped extensively~\cite{Sanna2017, Poggio2021, Hao2021} and numerous works were devoted to study its physical properties both in observations~\cite{Griv2017, Shen2020} and in numerical simulations~\cite{Khrapov2021, Vukcevic2022}. However, Milky Way is still only a single example of such a diverse and complex feature as spiral structure.

Overall, it is still puzzling which mechanisms exactly drive the formation of spiral structure, and there are multiple proposed scenarios. One of the most promising is the density wave theory~\cite{Lin1964, Shu2016} which suggest that spiral structure as a long-lived, quasi-stationary density wave in the disc of a galaxy. Another is dynamic spirals theory~\cite{Dobbs2014}, which aims to explain observed spiral structure as a combination of small short-living elements appearing from instabilities. It seems that there is no universal scenario of spiral arm formation, but instead different mechanisms may take place or even coexist in individual galaxies. It is not easy to determine which one is relevant in each specific case~\cite{Dobbs2018, Kostiuk2025}.

The distribution of light in spiral arms is one of their characteristics that was not studied exhaustively. Some studies deal with the morphology of spirals, which is essentially a qualitative description of the light distribution. Since~\cite{Elmegreen1987}, this framework has resulted in a commonly accepted classification of spiral structure into three types: grand-design (G), multi-armed (M) and flocculent (F). Considering the more quantitative approach,~\cite{Savchenko2020, Mosenkov2020, Mosenkov2024} are examples of studies in which the light distribution in the spiral arms was extracted with slices perpendicular to the spiral arm. In these works, various properties of spiral arms such as pitch angles and their variations, widths, and exponential scale of the radial brightness distribution were measured.

It is often discussed how different parts of this distribution can be connected to the fundamental properties of the spiral arm. In particular, corotation radius is a valuable instrument to determine whether spiral structure in a galaxy is produced by a density wave~\cite{Lin1964} or not. There is a number of methods to measure them, including a few methods that utilize the surface brightness distribution in spiral arm. Various morphological indicators are expected, such as the truncation of spiral arms at certain resonance radii~\cite{Elmegreen1992} or gap in spiral arms at corotation resonance~\cite{Pan2023}. Other methods are based on offsets between spiral arm locations seen in different bands~\cite{Abdeen2020}, or make use of the varying asymmetry of spiral arm profile~\cite{Marchuk2024a}. As such, the detailed examination of the light distribution in spiral arms is important, and may possibly shed light to the long-standing question of their nature and origin.

When it is needed to describe some property of spiral arm, both in theory and in observations, overly simple models are often used. In particular, logarithmic spiral is a commonly used approximation of the shape of spiral arms, however it is known that spiral arms often have varying pitch angle (i.e. angle between the tangent to the spiral arm and a perpendicular to a radius)~\cite{Kennicutt1981, Savchenko2013}, whereas in logarithmic spiral it is by definition constant. In~\cite{Font2019}, it was proposed to use a polynomial function in log-polar coordinates; sometimes, a broken-line sequence of logarithmic spirals is used to describe a continuous spiral arm~\cite{Honig2015,Diaz-Garcia2019}. Nevertheless, in all these works there was no goal to study what approach is the most optimal. Another known functional form to mention is a ``scaffold'' function~\cite{Ringermacher2009}, however it is only capable of producing symmetric two-armed spiral structure. Finally,~\cite{Peng2010} shown that a combination of Fourier and bending modes is capable of reproducing a complex brightness distribution in spiral arms, however the number of parameters in the implemented model is large and most of parameters are hard to interpret.

To the best of our knowledge, there were no attempts to examine the functional form of this distribution. For example, in mentioned~\cite{Savchenko2020} pitch angle variations in spiral arms were measured, but the question of how exactly pitch angle changes was not considered. Exponential scale of radial surface brightness distribution was measured, but it was not examined if exponential function fits the observed distribution well. In our works devoted to the decomposition of galaxies with spiral arms~\cite{Chugunov2024, Marchuk2024b, Chugunov2025}, we have developed and used the 2D photometric model of spiral arm, but the functional form of this model was not a result of a thorough analysis. Rather, it was based on a compilation of various known properties of spiral arms and assumptions that seemed plausible. However, these works gave us some clues on that some parts of the model need improvements, and some have to be reconsidered. In particular, radial brightness profiles of spiral arms may actually be far from exponential, despite spiral arms being a part of a disc, which is commonly described with the exponential profile, contrary to our initial assumption. We should also notice that we have no means to examine spiral structure in other galaxies in 3D like it can be done in our Milky Way~\cite{Griv2022, Wienen2022}, and we only discuss 2D light distribution in the galactic plane.

Considering all above, our ultimate goal for this study is to measure precisely the light distribution in spiral arms of galaxies, including more specific properties such as overall shape, light distribution along the spiral arm and brightness profile across it. Then, these results can be directly applied to construct a justified 2D photometric model of spiral arm for later use in decomposition, and to examine the connection between observed brightness distribution features and physical properties of galaxies and their spiral patterns.

\section{Data and methods}

\subsection{Images}

\begin{figure}[H]
\centering
\includegraphics[width=0.99\textwidth]{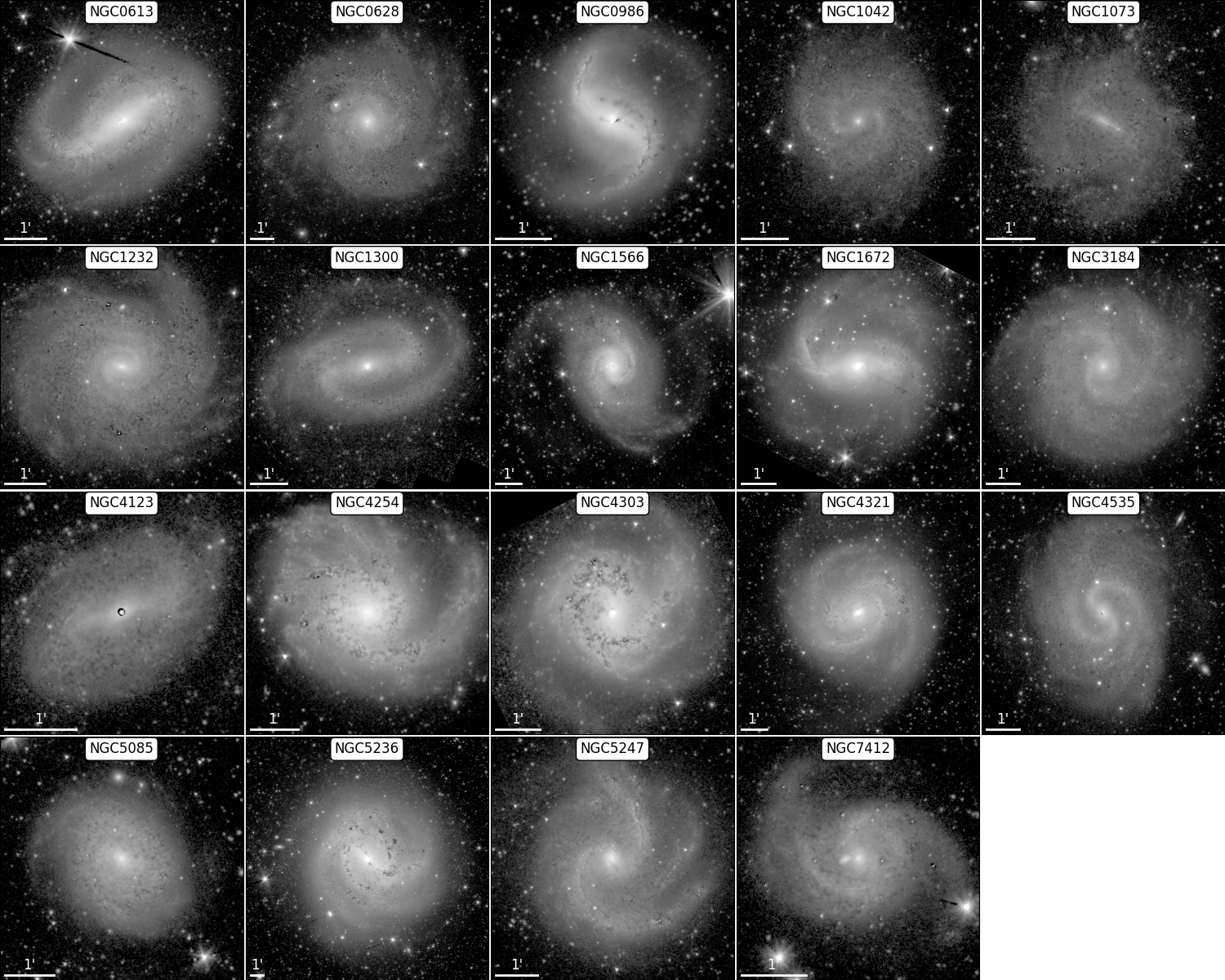}
\caption{Mosaic image showing all galaxies in the sample. Images in 3.6 $\mu$m corrected for non-stellar radiation are shown.\label{fig:mosaic}}
\end{figure}

In order to study the structure of spiral arms in detail, we select a sample of well-resolved, face-on galaxies with prominent spiral structure. We choose appropriate galaxies from S$^4$G survey~\cite{Sheth2010}. The original survey encompasses more than 2300 bright, nearby galaxies, imaged in 3.6 and 4.5 $\mu$m with a pixel size of 0.75 arcsec. As a primary selection criterion, we consider only galaxies that are angularly large ($R_{25.5} > 2'$), spiral galaxies ($T > 0$), seen nearly face-on ($i < 40^\circ$); all these parameters are taken according to parameters from this survey. We used images that were corrected for non-stellar emission in~\cite{Querejeta2015} using $[3.6]-[4.5]$ colors. As these corrected images are dominated by old stellar radiation, they can serve as stellar mass distribution maps, and therefore our study deals with the stellar mass distribution in spiral arms. Finally, we selected galaxies with prominent spiral structure by visual inspection. We have ended up with 19 objects, listed in Table~\ref{tab:sample} along with some basic parameters and shown in a mosaic image in Fig.~\ref{fig:mosaic}. Our sample consists of 13 multi-armed galaxies and 6 grand-design ones. Flocculent spirals are absent, because we initially selected objects with distinct, measurable elements of spiral structure, according to the main goal of our study.

\begin{table}[H]
\caption{Galaxies that are included in our sample.\label{tab:sample}}
		\begin{tabularx}{\textwidth}{CCCCC|CCCCC}
			\toprule
            Galaxy & $R_{25.5}$ (arc-min) & $\log M_*$ & $D$ (Mpc) & AC & 
            Galaxy & $R_{25.5}$ (arc-min) & $\log M_*$ & $D$ (Mpc) & AC\\
            \midrule
            NGC0613 & 3.28 & 11.1 & 25.1 & M &
            NGC0628 & 5.77 & 10.3 & 9.1 & M \\
            NGC0986 & 2.29 & 10.4 & 17.2 & G &
            NGC1042 & 2.66 & 9.6 & 9.4 & M \\
            NGC1073 & 2.53 & 10.0 & 15.2 & M &
            NGC1232 & 3.79 & 10.7 & 18.7 & M \\
            NGC1300 & 3.42 & 10.6 & 18.0 & G &
            NGC1566 & 4.39 & 10.6 & 12.2 & G \\
            NGC1672 & 3.92 & 10.7 & 14.5 & G &
            NGC3184 & 4.11 & 10.4 & 12.0 & M \\
            NGC4123 & 2.12 & 10.3 & 21.9 & M &
            NGC4254 & 3.20 & 10.7 & 15.4 & M \\
            NGC4303 & 3.78 & 10.9 & 16.5 & M &
            NGC4321 & 5.21 & 10.9 & 16.0 & G \\
            NGC4535 & 4.01 & 10.7 & 17.0 & M &
            NGC5085 & 2.33 & 10.8 & 28.9 & M \\
            NGC5236 & 9.48 & 11.0 & 7.0 & M &
            NGC5247 & 3.57 & 10.8 & 22.2 & G \\
            NGC7412 & 2.11 & 9.8 & 12.5 & M & & & & & \\
            \bottomrule
		\end{tabularx}
	\noindent{$R_{25.5}$ is a semi-major axis at $\mu(3.6 \mu\text{m}) = 25.5~\text{mag(AB)}/\text{arcsec}^2$, $\log M_*$ is a logarithm of total stellar mass of a galaxy (in solar masses), $D$ is redshift-independent distance measure, AC is arm class (G is for grand-design, M is for multi-armed). Data was taken from the original S$^4$G paper~\cite{Sheth2010}, except for arm class data from~\cite{Buta2015}.}
\end{table}

For the same sample of galaxies, we also made use of original 3.6 $\mu$m images from S$^4$G (uncorrected for non-stellar emission) and GALEX images in the FUV band (pivot wavelength is 1524~\AA~and pixel size is 1.5 arcsec) and performed, partially, the same analysis as for the main set of images. This allows us to examine not only stellar mass distribution, but also the ``natural'' light distribution in one band (original 3.6 $\mu$m) and the approximate star formation rate distribution, because FUV radiation is one of a direct tracers of star formation in timescale $\sim$100~Myr~\cite{Calzetti2013}.

\subsection{Extraction of the spiral arms}
\label{sec:extraction}
Automatic determination of the location of spiral arms in a galaxy is a difficult task (see e.g.~\cite{Davis2014, Forgan2018, Walmsley2023}). Instead, we rely on a method based on manual marking of spiral features, which itself is a commonly used technique~\cite{Masters2021}. Using SAOImage DS9~\cite{Joye2003}, we marked spiral features by placing multiple dots of some colour along a given arm, from the centre to the periphery. The coordinates of these points were further processed to minimise the human factor in tracing spiral arms (see below in this Section).

To examine the distribution of light in spiral arms, we utilize slightly modified slicing method from~\cite{Savchenko2020}, also similarly done in \cite{Querejeta2024}. Before applying slicing, we prepare images to extract only the spiral structure from the images. However, there is no widely accepted method of separating the spiral structure from the underlying disc. In order to ensure that our future results do not suffer from any bias introduced by our method of separation, we implement two different methods of extracting the spiral structure, and conduct most parts of our further analysis independently for two samples of images, obtained by different ways.

In a first method, we perform decomposition using IMFIT~\cite{Erwin2015} in an ordinary way. We use a model that includes bulge, disc, and a bar, when exists; disc parameters are used to determine the orientation of the galactic plane, whereas bulge and bar models are subtracted from the image. Then, to eliminate underlying disc contribution, we estimate its profile: for a given galactocentric radius $r$, we take all pixels at this radius, calculate quantile 0.1 and then subtract it. This method is motivated by that pixels with the lowest brightness at a given radius are free from spiral arm emission and can serve as an estimate of underlying disc. Further, we will refer to this method of subtraction and the corresponding sample of separated spiral structure images as ``$q = 0.1$ disc''.

For a second method of extracting spiral structure, we first construct the mask of spiral structure. Starting from the existing visual marking of spirals, we estimate the width of spiral arms at each point following the method that was used in~\cite{Reshetnikov2022, Reshetnikov2023}. Then, the mask is constructed as a polygon based on the inner and outer edges of spiral arm. Next, we perform the decomposition again: we use the same model as in previous method, but we also apply the mask described above. Therefore, spiral arm areas are already excluded from fitting, and the disc model obtained from decomposition this way can be treated as the estimate of underlying disc. Then, the entire decomposition model, including the disc, is subtracted, and the orientation of the galactic plane can be different compared to the $q = 0.1$ disc method. We will refer to this method and the corresponding sample of images as to ``decomposition disc''.

We observe that the decomposition disc is usually slightly brighter than $q = 0.1$ disc, likely not because the difference between methods themselves, but because of the numerical value of the threshold. We again note that there are no obvious reasons to adopt some exact method and exact threshold to separate spiral arm flux from the underlying disc as the most correct, and we employ both to test the robustness of our future results.

Next, based on how spiral arms were marked, a dense set of slices, each going across the spiral arm, is prepared. Each slice is radial, i.e. it follows the line of constant azimuthal angle $\varphi$, and each goes through a given spiral arm, not overlapping with others. Each slice is fitted with a Gaussian function, and these fit results serve as a preliminary estimation of the radius $r$ and the width $w$ of spiral arm at a given location. These results were also examined visually; in cases when fitted parameters were clearly inconsistent with the visually estimated position or width of the arm, these slices were removed. These inconsistencies are inevitable as small-scale features are abundant in spiral arms, and these removed slices are the reason of the non-uniform distribution of slices on the spiral arm e.g. in Figure~\ref{fig:str_arms_example} and~\ref{fig:NGC5247_shapes}.

\subsection{Straightening of the spiral arms}
\label{sec:straightening}
For the analysis following the determination of the shape of spiral arms (specifically, the ridge-line, defined in Section~\ref{sec:shapes}), we have performed some additional manipulations with images before doing the next steps.

It is natural to study spiral arms in polar coordinates, but since some arms can sweep multiple revolutions around a galactic center, it introduces an ambiguity into the measurements: some distinct arm locations would have the same values of $\phi$. To solve this problem, we introduce a new coordinate system $[\rho, \psi]$ in which an arm appears as a straight line. In this coordinate system the value of $\psi$ is an angular distance of some point on a spiral measured along the arm (so taking into account multiple revolutions) from its beginning as seen from the galactic centre. The value of $\psi$ is not, therefore, limited between 0 and $2\pi$, but can take higher values reflecting the total arm's length. The second coordinate $\rho$ is effectively a coordinate across a spiral arm. It is measured as a difference between the galactocentric distance of some point $r$ and the ridge-line of the arm $r(\psi)$ that is described by some smooth analytical function (see Section~\ref{sec:shapes}.

\begin{figure}[H]
\centering
\includegraphics[width=0.99\textwidth]{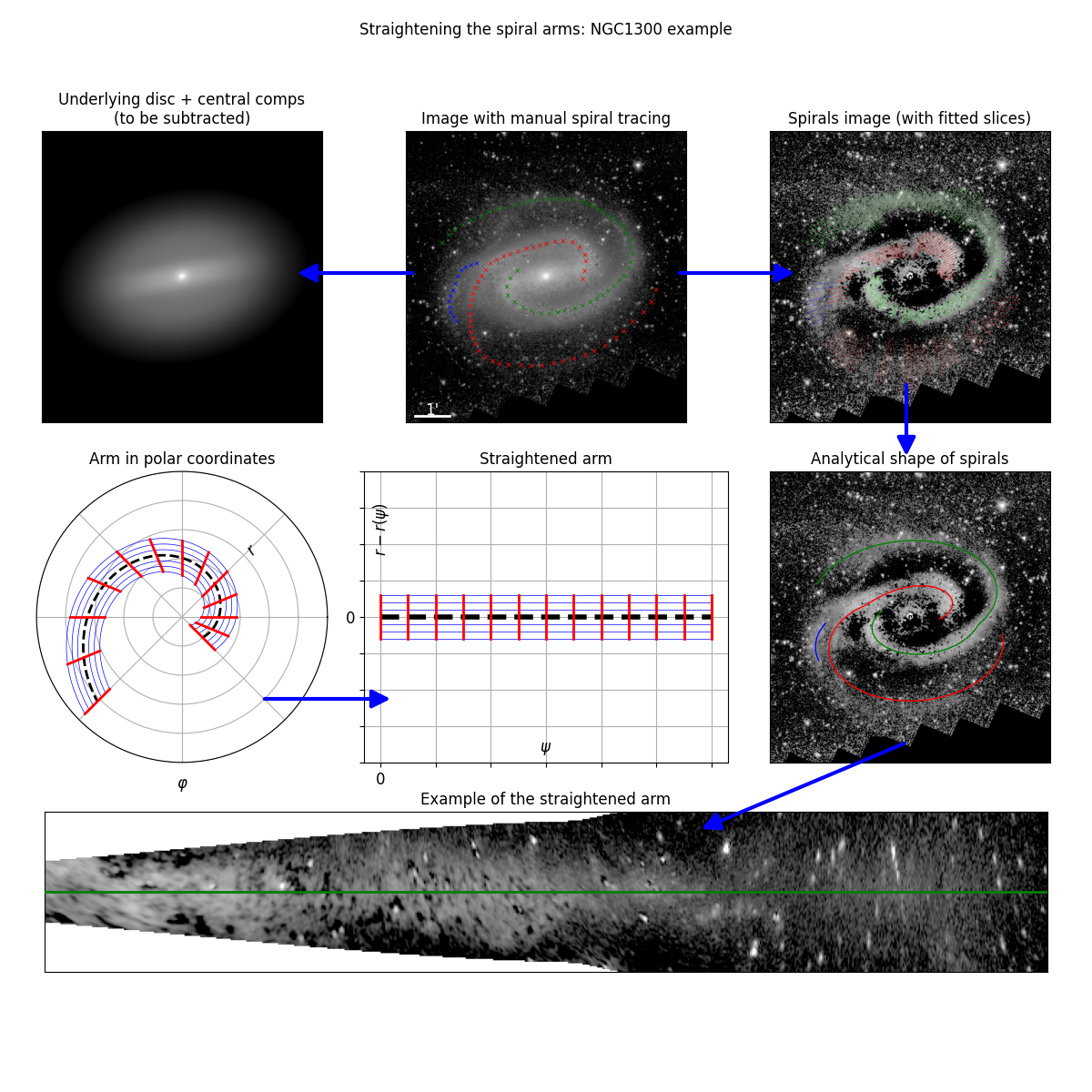}
\caption{A schematic representation of the image processing, with NGC~1300 shown as an example. We start from the original images from~\cite{Querejeta2015} (top middle); central components and underlying disc (top left) are estimated and subtracted to obtain spiral arms image (top right). Based on manual tracing of spiral arms, slices are prepared and an analytical approximation of the spiral arm shape is calculated (middle right). These analytical approximations are used to construct straightened spiral arm images (bottom). The essence of straightening process is shown schematically (middle left and centre).\label{fig:str_arms_example}}
\end{figure}

Images of each individual spiral arm in these coordinates for the corresponding $r(\psi)$ were prepared; we call them images of straightened spiral arms. In Figure~\ref{fig:str_arms_example}, we summarise our manipulations with the image, required for further analysis, including spiral arms straightening. Finally, using straightened spiral arms, we can examine precise distributions of light from radius and from azimuthal angle, putting aside the question of shapes and considering spiral arms separately from each other. Also, we construct a set of straightened spiral arms in $[r, \psi - \psi(r)]$ coordinates, where $\psi(r)$ is the inverse of $r(\psi)$; in this case, the second coordinate is the azimuthal distance from the point to the spiral arm. As an example, this can be used to analyse azimuthal offsets between UV and IR images, which is useful for finding corotation radii~\cite{Kostiuk2024}.

\section{Spiral arms measurements}

In this Section, we first present the main results of the application of our method and describe photometric properties of spiral arms. Then, we propose a photometric model of spiral arm summarizing these properties accurately, and check how well it represents real spirals during the decomposition.

Overall, we have measured 88 spiral features, including large spiral arms as well as small spurs, in 19 galaxies. To formalize distinction between large and small spiral features, we measure their azimuthal length $l_\psi$, which is the value of the $\psi$ at the end of the arm. Here and after, we call features with $l_\psi < 90^\circ$ spurs and refer to others as to spiral arms. Note that the term ``spurs'' is often used specifically to denote small-scale features which jut out from major spiral arms, and there are some different terms for such small-scale features~\cite{Weaver1970, Elmegreen1980, Perez-Villegas2015}, but in this paper, we use term ``spurs'' for all spiral-like features with $l_\psi < 90^\circ$, regardless of their location and other properties. In overall, out of 88 spiral features 26 are spurs and 62 are spiral arms, from which 35 have $l_\psi \geq 180^\circ$. Out of 62 spiral arms, 17 belong to 6 grand-design galaxies and the remaining 45 are found in 13 multi-armed galaxies. Interestingly, the difference in average azimuthal length of spiral arms between two types is smaller than one could have expect: for grand-design galaxies, it is $273^\circ$, and for multi-armed it is $244^\circ$. Given the standard deviations ($143^\circ$ and $131^\circ$) and limited size of the sample, this difference is not statistically significant. In particular, three longest spiral arms in the entire sample with $l_\psi > 540^\circ$ are found in NGC~628 and NGC~1232, both are classified as multi-armed, according to~\cite{Buta2015}.

As an example, in NGC~1300 (Figure~\ref{fig:str_arms_example}) there are 2 spiral arms (marked green and red) and 1 spur (marked blue), and in NGC~5247 (Figure~\ref{fig:NGC5247_shapes}) there are 3 spiral arms (blue, green, red) and 1 spur (cyan).

\subsection{Shapes of spiral arms}
\label{sec:shapes}

We define the shape of spiral arm as its ridge-line $r(\psi)$ in polar coordinates, described above (Section~\ref{sec:straightening}). The ridge-line a curve that, for a given $\psi$, yields $r$ where the brightness is highest.

For each spiral feature, we have multiple slices at different locations. For each slice, we determine its peak location $[r, \psi]$ by fitting a Gaussian to the brightness distribution along the slice (see Section~\ref{sec:extraction} and Figures 5, 6, 8 in~\cite{Savchenko2020}). This way, we obtain multiple points tracing the ridge-line of the spiral arm, referring to this set of points as observed $r(\psi)$. Next, we aim to select the analytical function that fits observed $r(\psi)$ good enough.

The average pitch angle of spirals in our sample is $\mu = 18.9^\circ$, in agreement with literature (e.g.~\cite{Yu2018a, Diaz-Garcia2019} with values of $20.5^\circ$ and $19.6^\circ$, respectively, obtained with Fourier methods). If one considers only long spiral arms ($l_\psi > 180^\circ$), then average pitch angle decreases to $17.5^\circ$. It is noteworthy that some works focused on measuring pitch angles of spiral arms using slicing or decomposition. In these works the results are usually smaller: in particular,~\cite{Savchenko2020, Chugunov2024} present average pitch angles of $14.8^\circ$ and $15.9^\circ$, respectively. This discrepancy probably can attributed to the fact that shorter spiral arms are more difficult to trace and therefore are more likely to be missed in such kind of a study. In the same time, they can have larger pitch angles, on average, as seen in~\cite{Chugunov2025}. Interestingly, in~\cite{Yu2020} the bimodality of the pitch angle distribution is reported, with peaks near $12^\circ$ and $23^\circ$. However, we cannot check the possible bimodality, as our sample is not nearly as large as theirs, consisting of more than 4300 galaxies. Finally, the average pitch angle values of spiral arms in grand-design and multi-armed galaxies are almost the same, being $\mu = 19.1^\circ$ and $\mu = 18.8^\circ$, respectively.

\subsubsection{Functions overview}
\label{sec:functions_overview}
\begin{figure}[H]
\centering
\includegraphics[width=0.99\textwidth]{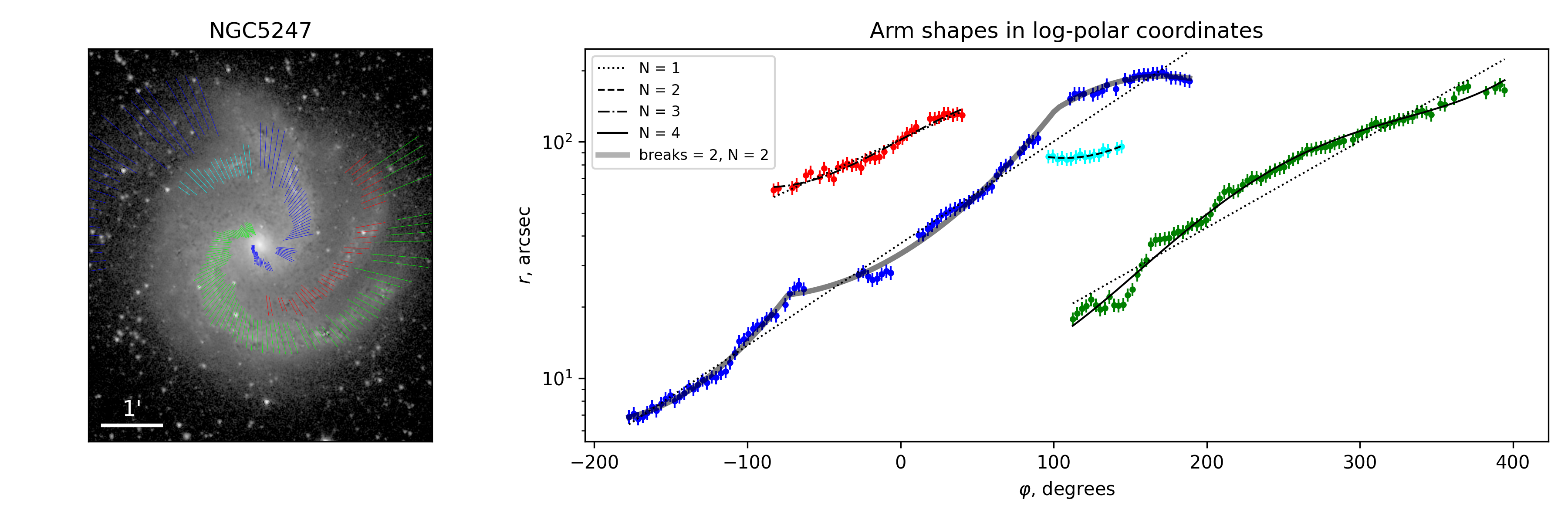}
\caption{Image of NGC~5247, where coloured lines mark the location and width of spiral arm (left); $r(\psi)$ plot for all spiral arms with different functions used to fit the data. Reproduced from~\cite{Chugunov2025_Kourovka}.\label{fig:NGC5247_shapes}}
\end{figure}

The most commonly used function to describe $r(\psi)$ is the logarithmic spiral. Its pitch angle is constant and it is used both in theoretical and observational works thanks to its simplicity. However, it is known that pitch angles in observed spiral arms are not constant~\cite{Savchenko2013}. In our work, we also observe a high degree of variations in pitch angle, most clearly seen in $\log r$--$\psi$ plot (an example is Figure~\ref{fig:NGC5247_shapes}, discussed further on), where logarithmic spiral appears as a straight line.

It was proposed in~\cite{Font2019} that $[\psi, \ln r]$ may be better fitted by polynomial rather than linear function, which is equivalent to the following Function~\ref{eq:r_psi}. One can see that logarithmic spiral is a special case when $N = 1$, and therefore we call our function polynomial-logarithmic spiral. We start with this function to fit $r(\psi)$.

\begin{linenomath}
\begin{equation}
	\label{eq:r_psi}
	r(\psi) = r_0 \times \exp \left(\sum_{n=1}^N k_n \psi^n\right)\,.
\end{equation}
\end{linenomath}

This function can be easily extended to high enough $N$ to fit more complex shapes of spirals. Another advantage is that pitch angle $\mu$ varies with $\psi$ simply as a polynomial of $N - 1$ degree. In our previous works devoted to decomposition, we adopted $N = 4$ for our shape function. However, one should use higher-order polynomials with caution, because they have, at best, limited physical meaning and they are difficult to control in fitting.

Visual inspection of images leads to a conclusion that spiral arms often have bendings, i.e. locations where the pitch angle changes abruptly. Again, in~\cite{Font2019} it was mentioned that ``breaks'' in pitch angles of spiral arms exist, and they are not mutually exclusive with gradual variation of pitch angle along the spiral arms. The canonical example for this is M~51~\cite{Font2024}: both its spiral arms have regions of smoothly varying pitch angle, as well as abrupt bendings. Therefore, at least in some galaxies spiral arms cannot be described with any smooth function like Function~\ref{eq:r_psi} very good. To account for these cases, we can simply use a piecewise function consisting of two or three smooth parts, each described by Function~\ref{eq:r_psi} with the independent set of $k_n$ coefficients, which is represented in Function~\ref{eq:r_psi_bends}. We used this approach in our decomposition study of M~51~\cite{Marchuk2024b}.

\begin{equation}
\label{eq:r_psi_bends}
r(\psi) =
\left\{\begin{array}{ll}
    r_0 \times \exp \left(\sum_{n=1}^N k_{n, 0} \psi^n\right), & 0 \leq \psi < \psi_1\,\\
    \ldots \\
    r_m \times \exp \left(\sum_{n=1}^N k_{n, m} (\psi - \psi_m)^n\right), & \psi_m \leq \psi\,.
\end{array}\right.\
\end{equation}

In practice, we find that no more than 2 bends are needed, therefore $m \leq 2$ and there are no more than 3 independently defined polynomials. Note that $r_m$ parameters are not independent except $r_0$, because spiral arm is continuous and the next segment starts at the same radius as previous ends.

\subsubsection{Comparison of the functions}

For each slice, there is a some degree of uncertainty $\sigma_r$ in $r$, manifesting itself in a noticeable scatter of dots (see Figure~\ref{fig:NGC5247_shapes}). To get an understanding of the scale of this uncertainty, we compare it to measured widths (specifically, FWHM) of spiral arms $w$. We observe that the root mean square of $\sigma_r / w$ in arms in our sample is, on average, $0.16 \pm 0.06$.

When one fits a function to the observed $r(\psi)$, some residual $\delta_r$ remains. Then, assessing fit quality, we can use $\chi^2$ statistic (and use it in a conventional way when dealing with BIC later in this Section). However, to keep in touch with the characteristic scale of spiral structure, here we will present another parameter, namely the root mean square of $\delta_r / w$, to characterise the goodness of fit. The motivation underlying this choice is connected with the way how decomposition with spiral arms is conducted in practice. To fit an observed spiral arm, each slice of the model arm can have displacement from the observed arm much smaller than its width.

\begin{figure}[H]
\centering
\includegraphics[width=0.99\textwidth]{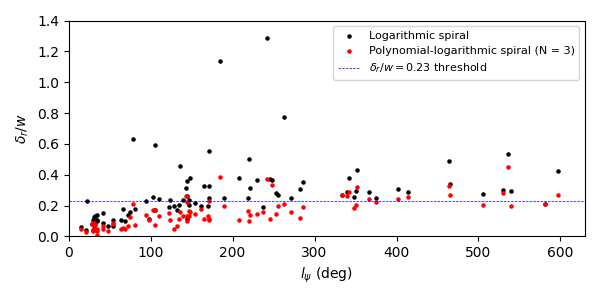}
\caption{Each dot correspond to a single spiral feature, showing $\delta_r$ and $l_\psi$ for all spirals in our sample. Black dots represent $\delta_r$ for logarithmic spiral fits, whereas red dots correspond to fit with the polynomial-logarithmic spiral of $N = 3$. Dashed line show the threshold of $\delta_r / w = 0.23$\label{fig:deltaw_lpsi}}
\end{figure}

As we described, $\sigma_r / w = 0.16$ on average, and we should not expect that $\delta_r / w$ will be lower than this, even when fit is considered to be good. Also, if $\delta_r / w$ is $\sqrt{2} \times 0.16 = 0.23$, then the contribution of some systematic offset between the model and observed $r(\psi)$ is roughly equal to the contribution from the $\sigma_r$ (statistically, it is the case of $\chi^2 = 2$). Now, if we consider $0.23 w$ as a threshold, then we observe, as expected, that logarithmic spiral is usually unable to follow the shape of real spiral arms well; among spiral arms, only in 23\% cases $\delta_r / w$ is smaller than this threshold. However, for spurs this fraction is increased to 92\% which is expected since pitch angle is unlikely to change significantly along the short spur. In Figure~\ref{fig:deltaw_lpsi}, we show the diagram of $\delta_r$ versus $l_\psi$ for all spirals in our sample, which $r(\psi)$ was fitted with logarithmic spiral and with polynomial-logarithmic spiral of $N = 3$.

As we test functions with a different number of parameters, we expect those with higher number of parameters to perform better in terms of average deviation. But instead, our goal is to select a function with the optimal number of parameters, which does not introduce overfitting, and therefore we use BIC statistic (Bayesian Information Criterion,~\cite{Bailer-Jones2017}). For the case of the random Gaussian noise, it essentially adds a penalty to $\chi^2$ statistic depending on the number of data points $N$ and the number of free parameters $k$. In its simplest form, BIC is described by Equation~\ref{eq:BIC}; in more complex cases, BIC is modified accordingly, in particular in the case when not all data points are independent, for example, due to PSF~\cite{Simard2011, Head2014}.

\begin{linenomath}
\begin{equation}
	\label{eq:BIC}
	\text{BIC} = \chi^2 + k \ln N.
\end{equation}
\end{linenomath}

A model which is too complicated, i.e. has too many free parameters, will have higher (worse) BIC statistic than a model with lower number of free parameters despite having somewhat better $\chi^2$. Therefore, if the addition of a new parameter increases the BIC of a model, it means that overfitting appears. However, if some complex model yields the lowest BIC, it does not mean that one has necessarily use it instead of some more simple model.

We fit functions~\ref{eq:r_psi} with different $N$ and~\ref{eq:r_psi_bends} with 1 or 2 bends to the each single arm. We measure BIC of these fits; some function yields the smallest BIC, and if some other function has higher number of parameters and significantly higher BIC, we interpret it as a sign of overfitting. Following~\cite{Raftery1995}, we only consider BIC difference more than 2 as significant, i.e. if more complex function yields higher BIC but the difference is smaller than 2, it is not considered overfitting. Naturally, the longer spiral arm is, the higher number of parameters may be justified, and we draw different conclusions for spirals of different length, described below:

\begin{itemize}
    \item For 26 spurs ($l_\psi < 90^\circ$), 2 or 3 parameters are optimal and do not introduce overfitting in all cases (corresponding to Equation~\ref{eq:r_psi} with $N = 1$ or $N = 2$). If the number of parameters is increased to 4, it yields overfitting in more than a half cases.
    \item For 27 relatively short spiral arms ($90^\circ \leq l_\psi < 180^\circ$), 4 parameters (Equation~\ref{eq:r_psi} with $N = 3$) is optimal in most cases; overfitting is present only for 3 spiral arms. Increasing the number of parameters leads to overfitting in more than a half cases.
    \item For 23 relatively long spiral arms ($180^\circ \leq l_\psi < 360^\circ$), 6 parameters (Equation~\ref{eq:r_psi} with $N = 5$, or Equation~\ref{eq:r_psi_bends} with 1 bending and $N = 2$, or with 2 bendings and $N = 1$) are optimal in most cases; overfitting is present only for 3 spiral arms. With 8 parameters (Equation~\ref{eq:r_psi_bends} with 1 bending and $N = 3$), 9 arms are overfitted.
    \item For 12 longest spiral arms ($l_\psi \geq 360^\circ$), no overfitting appears at least up to 8 parameters.
\end{itemize}

This analysis helps us to determine the most complex functions that can be used as ``default'' ones, but it does not mean that the most complex possible function will be used as default. In particular, bendings should be included only when they are observed in a real arm, and, higher-order polynomials probably should not be used unless needed, for the reasons mentioned above in Section~\ref{sec:functions_overview}. On the other hand, in specific cases even more complex functions can be used, as there is a fraction of galaxies where it is needed to achieve higher quality of fit (see below), and it can be done without overfitting.  

If we consider some function, for example the polynomial-logarithmic spiral with $N = 3$ or pure logarithmic spiral, we observe that the average $\delta_r / w$ typically depends on azimuthal length of spiral arm (see Figure~\ref{fig:deltaw_lpsi}). As one should have expected, the average deviation for the shortest spurs is small and it increases towards longer spiral arms. The natural explanation of this trend is that pitch angle is unlikely to vary strongly along the extent of a small spur. Which is more interesting, however, is that this behaviour is not monotonous. Over the range of approximately $100^\circ \leq l_\psi < 300^\circ$, there are a few spiral arms with exceptionally high $\delta_r / w$: in this range, 5 arms from total of 42 have $\delta_r / w > 0.55$, but none of 18 spirals with $l_\psi \geq 300^\circ$ have $\delta_r / w$ this high. In other words, spirals of moderate length have highest deviation from the logarithmic spiral, compared to both spurs and the longest spiral arms. This can be interpreted as a consistency with~\cite{Savchenko2020}: authors found that pitch angle variation of spiral arms for multi-armed galaxies is higher than both for flocculent and grand-design spiral galaxies. Note that pitch angle variation represents the difference from the logarithmic spiral to some degree, and the different types of spiral structure mentioned above correspond to spirals with different lengths. In flocculent galaxies, spiral structure consists of numerous short spurs; multi-armed galaxies host a few conspicuous spiral arms of moderate lengths; grand-design galaxies have a pair of long, symmetric spiral arms~\cite{Elmegreen1987}. Considering this, our finding is in agreement with~\cite{Savchenko2020}. Also, in~\cite{Chugunov2024} we have found that galaxies with relatively brighter spiral arms have smaller pitch angle variations. Considering that grand-design galaxies have brighter spiral structure than multi-armed~\cite{Savchenko2020}, it also aligns with this finding.

Finally, we combine our results concerning the applicability of the different functions for describing the shapes of spiral arms:

\begin{itemize}
\item As mentioned above, most spurs are already well-described by logarithmic spirals, but using a polynomial-logarithmic spiral with $N = 2$ yields $\delta_r/w < 0.23$ in all cases, without exceptions.
\item We observe that spiral arms ($l_\psi > 90^\circ$) are described by the polynomial-logarithmic spiral with $N = 3$ much better than by logarithmic spiral. In 73\% cases, they have $\delta_r/w$ smaller than the threshold of 0.23 discussed above. Among the short spiral arms ($90^\circ \leq l_\psi < 180^\circ$), this fraction is 93\%.
\item For longer spiral arms ($180^\circ \leq l_\psi < 360^\circ$), polynomial-logarithmic spirals with $N = 3$ and $N = 4$ yield $\delta_r/w < 0.23$ in 70\% and 83\% cases, respectively. For the remaining cases, often bendings are the reason why $\delta_r/w$ is still larger than this threshold; using polynomial-logarithmic spiral with $N = 4$ or either Function~\ref{eq:r_psi_bends} with $m = 1$ and $N = 2$ (which does not introduce overfitting for this range of azimuthal lengths), one can achieve $\delta_r/w < 0.23$ in 96\% cases.
\item For the longest spiral arms ($l_\psi > 360^\circ$), polynomial-logarithmic spiral with $N = 4$ yields $\delta_r/w < 0.23$ in 67\% cases, and again, bendings are playing role in this. Remaining cases, however, can be fitted with the Function~\ref{eq:r_psi_bends} with $m = 1$ and $N = 3$, yielding $\delta_r/w < 0.23$ without exceptions.
\item In overall, we find 35\% of spiral arms exhibit bendings, which was determined after the careful inspection of images, $r(\psi)$ diagrams and comparison of models that include bendings versus models which do not. In grand-design galaxies, 47\% of spiral arms possess bendings, whereas in multi-armed galaxies, 31\% of spiral arms are bent. But, according to the two-proportion Z-test, there is not enough evidence to consider this difference to be significant. Bending locations are often associated with bifurcations of spirals or the passages near ends of the bar. In some cases, bendings are weak and simple polynomial-logarithmic function is enough to fit the spiral arm.
\end{itemize}

Concerning all above, we conclude that polynomial-logarithmic functions with $N = 2, 3, 4$ can be treated as ``default'' models, with the exact choice of $N$ depending on the length of a spiral: $N = 2$ for spurs with $l_\psi < 90^\circ$, $N = 3$ for spiral arms with $90^\circ \leq l_\psi < 180^\circ$ and $N = 4$ for longer spiral arms. However, for spirals with $l_\psi \geq 180^\circ$ bendings become common, and if any smooth function fails to fit the $r(\psi)$ of the spiral arm along its extent, one should consider using one or two bendings to describe this function.

\subsubsection{Testing alternative functions}

As mentioned in~\cite{Savchenko2013}, there are two more simple functions sometimes used to describe the $r(\psi)$ of spiral arms, other than logarithmic spiral. Specifically, there are Archimedean spiral (Equation~\ref{eq:archimedean}) and hyperbolic spiral (Equation~\ref{eq:hyperbolic}):

\begin{linenomath}
\begin{equation}
	\label{eq:archimedean}
	r(\psi) = r_0 + k \psi\,,
\end{equation}
\end{linenomath}

\begin{linenomath}
\begin{equation}
	\label{eq:hyperbolic}
	r(\psi) = a / (\psi - \psi_0)\,.
\end{equation}
\end{linenomath}

We also try to generalise these functions in the same manner as it was done by Equation~\ref{eq:r_psi} from logarithmic spiral, i.e. replacing simple $\psi$ with a polynomial $p(\psi)$. In our case, we have chosen a polynomial of $N = 3$ degree, and we will refer to these generalised models as polynomial-Archimedean and polynomial-hyperbolic spirals. Also, visual examination of $r(\psi)$ may suggest that shape function can be described as a some periodic wave superimposed on a logarithmic spiral (see Figure~\ref{fig:NGC5247_shapes}), and we also test a ``wave spiral'' function (Equation~\ref{eq:wave}) with the same number of free parameters as polynomial-logarithmic spiral with $N = 4$ has:

\begin{linenomath}
\begin{equation}
	\label{eq:wave}
	r(\psi) = r_0 \times \exp (k \psi) \times [1 + A \sin(\nu \psi + p)]\,.
\end{equation}
\end{linenomath}

Using the simple Archimedean and hyperbolic spirals to fit $r(\psi)$, we ensure that, on average, both of these functions cannot fit observed $r(\psi)$ of spiral arms significantly better than logarithmic simple spiral does. For spiral arms ($l_\psi > 90^\circ$), we achieve good fit ($\delta_r/w < 0.23$) in 23\% of cases for logarithmic spirals, 27\% of cases for Archimedean spirals and in 15\% of cases for hyperbolic. Mean $\delta_r/w$ values are 0.34, 0.42 and 1.22 for logarithmic, Archimedean and hyperbolic spirals, respectively. Concerning more complex functions, we observe that, given the same number of free parameters, the polynomial-Archimedean spiral yields approximations of $r(\psi)$ as well as polynomial-logarithmic spiral (Equation~\ref{eq:r_psi}), but polynomial-hyperbolic fails to produce a good fit in most cases. If we consider $N = 3$, the fraction of spiral arms fitted with $\delta_r/w < 0.23$ is 73\% for polynomial-logarithmic spiral, 74\% for polynomial-Archimedean spiral and 27\% for polynomial-hyperbolic. Wave spiral function, despite having number of parameters higher by 1, yields good fit quality only in 71\% cases. The main possible reason for these results is that hyperbolic spiral reaches infinite $r$ at a certain $\psi$, making it more difficult to control, especially when a generalization with polynomials is undertaken. Also, it is probably connected to the fact that spiral arm pitch angles tend to decrease from the beginning to the end, which is true for Archimedean spiral, and possible in wave spiral, but for hyperbolic the opposite is true. Specifically, we observe that 57\% of spirals have decreasing pitch angles towards their ends, and this proportion remains the same if only long spiral arms ($l_\psi > 180^\circ$) are considered. Our result is comparable with value of 64\% in~\cite{Savchenko2013} and significantly smaller than 80\% in~\cite{Chugunov2025}. The inconsistency with the latter work is most likely caused by the fact that it dealt with galaxies at large $z$. In~\cite{Diaz-Garcia2019}, their results (Tables 3 and 4) are interpreted as a lack of significant changes of pitch angles with distance, although the innermost part of spiral structure in their data (inside 1 disc exponential scale $h$ or 0.2 $R_{25}$) has pitch angles by a few degrees larger than the remaining parts.

\subsection{Distribution of light along spiral arms}
\label{sec:I_parallel}
Dealing with straightened spiral arms (Section~\ref{sec:straightening}), we can examine how surface brightness changes along each individual spiral arm. To do this, we fit a Gaussian function to each slice across the straightened spiral arm, centred at the ridge-line, and consider the amplitude of the fitted Gaussian to be spiral arm brightness $I$ at $[r(\psi), \psi]$. Then, we can analyse $I$ as a function of $r$, $\phi$ or both parameters.

\begin{figure}[H]
\centering
\includegraphics[width=0.99\textwidth]{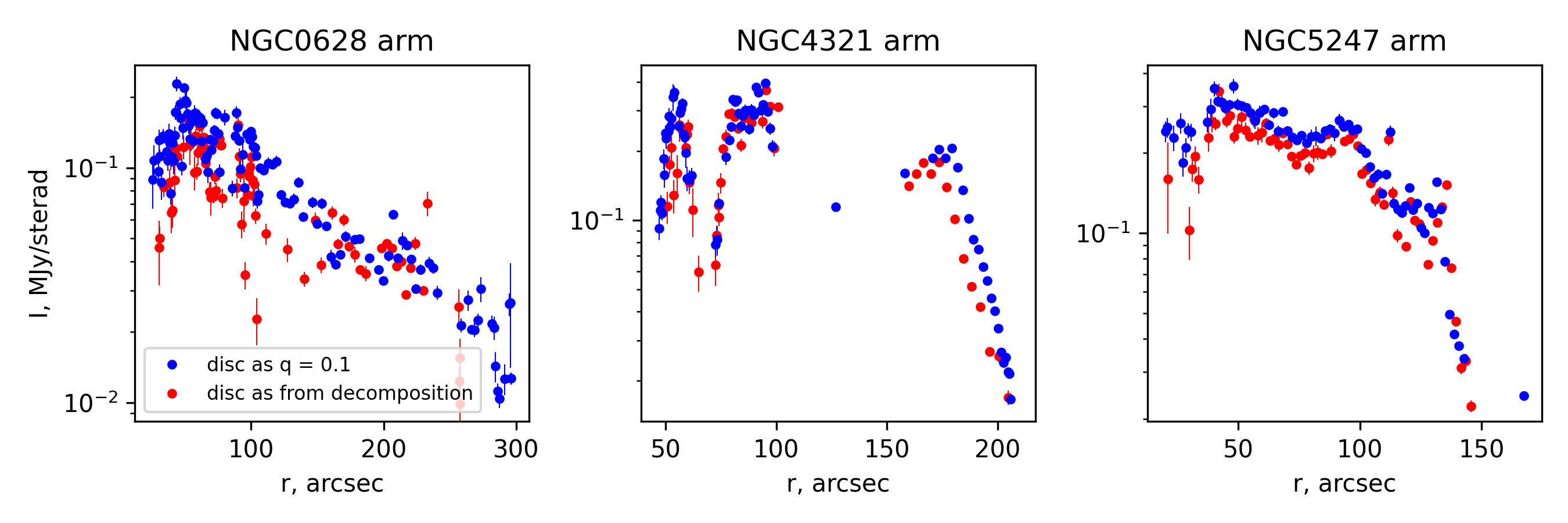}
\caption{Infrared surface brightness radial profiles of spiral arms for three examples: single arm in NGC~628, NGC~4321 and NGC~5247. Dots of different colors represent different methods of disc subtraction (Section~\ref{sec:extraction}): blue is $q = 0.1$ disc, red is a decomposition disc. Reproduced from~\cite{Chugunov2025_Kourovka}.\label{fig:profiles_examples}}
\end{figure}

The most striking feature we observe is that radial profile $I(r)$ is far from exponential in many cases, and differs significantly from arm to arm in overall (see examples in Figure~\ref{fig:profiles_examples}). In particular, both main arms in NGC~628 have nearly exponential radial profile. Spiral arm in NGC~5247 gives an example of a radial profile which decreases more or less monotonously, but the radial scale of brightness decrease is not constant at different radii. Finally, spiral arm in NGC~4321 is an extreme example of non-monotonous profile, having two ``dips'' in brightness distribution and an abrupt truncation at the end.

We see that our results do not depend of the disc subtraction method (Section~\ref{sec:extraction}). For both $q = 0.1$ disc and for decomposition disc, the overall spiral arm brightness profile remains the same, including the presence of surface brightness dips and other features (see Figure~\ref{fig:profiles_examples}). Finally, we observe at least one surface brightness dip, visible in images or in $I(r)$ diagram, in 24\% of spiral arms. In spiral arms of grand-design galaxies, there is a higher occurrence of dips (35\%) than in multi-armed ones (20\%), but, again, according to the two-proportion Z-test, the samples are too small (there are 17 arms in grand-design galaxies and 45 in multi-armed) to consider this difference to be statistically significant (p-value is 0.21).

In~\cite{Chugunov2024, Marchuk2024b, Chugunov2025}, various properties of spiral arms were measured using decomposition. Initially, there was an assumption that exponential decrease is a good approximation for spiral arm profile, similarly to the disc which host spiral arms. Therefore, the model of spiral arms was designed to produce spiral arms with exponential radial profile at the most part of the arm, modified to make smooth transition to zero near the ends of spiral arm (growth and cutoff). It was intended that growth and cutoff parts are small and needed only to make spiral arms have finite extent and have smooth truncation. However, in these works it was found that measured exponential scales of spiral arms differ significantly and are poorly consistent for arms in a single galaxy or for a single arm in close but different photometric bands. Considering all above, in~\cite{Chugunov2025} it was proposed that many spiral arms have non-exponential radial profiles, which leads to that exponential scales being determined poorly and, actually, have limited physical meaning at best. Despite that, decomposition method in these works usually was able to reproduce the brightness profile of spiral arms well, likely because growth and cutoff parts could be fitted to a large values, thus making spiral arm non-exponential over the significant part of it. Now, when spiral arm brightness profiles are measured directly, this conjecture is confirmed.

Although the existence of growth and cutoff parts is now demonstrated, one cannot be completely sure which function fits them best. There were different modifications of our function in~\cite{Chugunov2024, Marchuk2024b, Chugunov2025}, but in all cases they were arbitrary. We are going to test different possible functions possible by fitting 2D models to the images of straightened arms (see Section~\ref{sec:2d_fitting}).

\subsection{Widths of spiral arms}
\label{sec:widths}
As in the previous section, we deal with straightened spiral arms (Section~\ref{sec:straightening}). Fitting Gaussian functions into slices, we measure the widths of spiral arms at different points of spiral arms. We consider FWHM as a measure of spiral arm width. One should note that we actually measure a radial width $w_r$, but for the case of typical pitch angles of spiral arms $\mu$, it differs a little from the true width $w$, measured perpendicularly to the spiral arm (they differ by the factor of $\cos \mu \approx 0.94$ for $\mu = 20^\circ$)

\begin{figure}[H]
\centering
\includegraphics[width=0.99\textwidth]{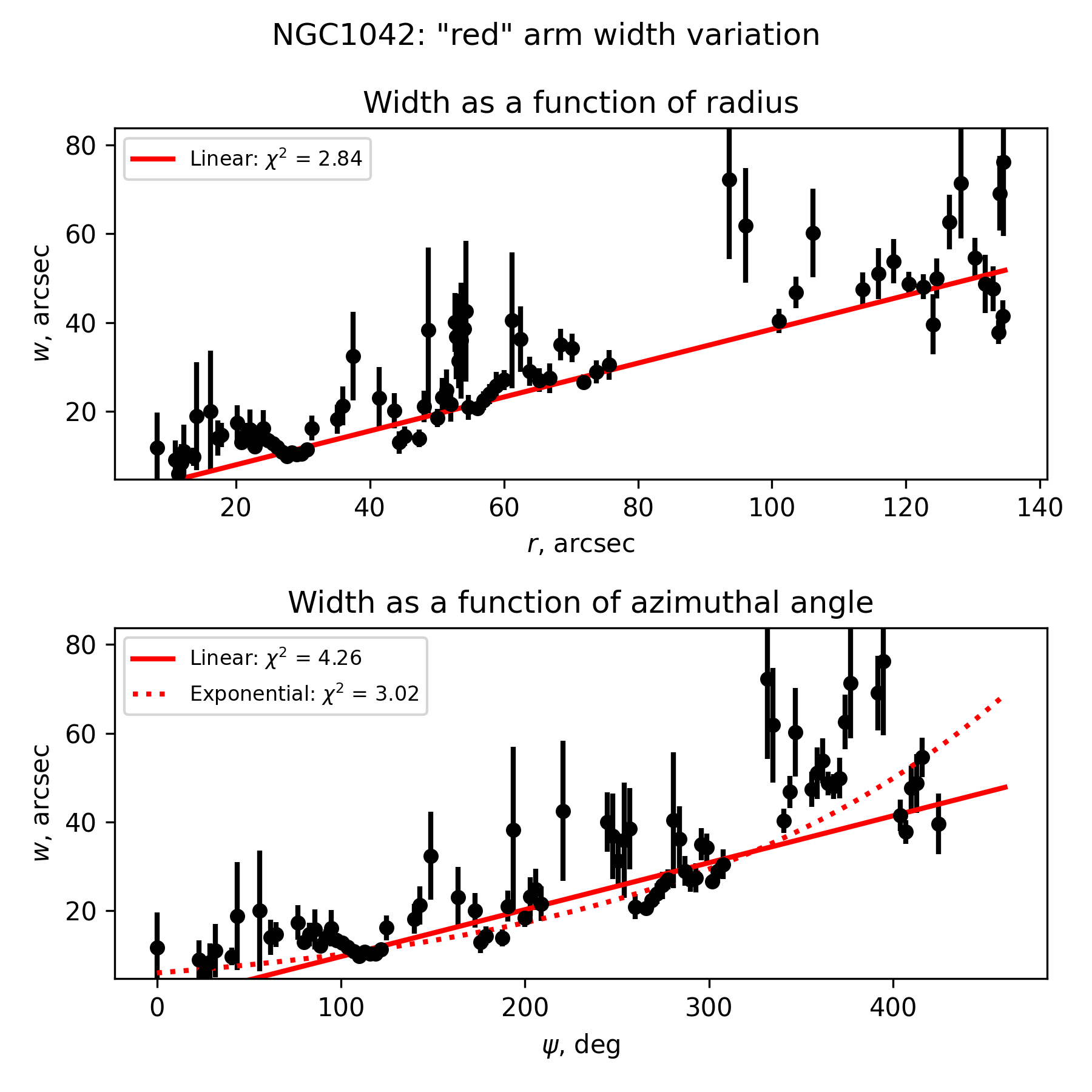}
\caption{An example of width variation in a spiral arm of NGC~1042. Each point is a single width measurement at some point; on top, measurements are presented as a function of radius $r$, and on bottom, as a function of azimuthal angle $\psi$. Simple functions are fitted, and fit quality is shown.\label{fig:NGC1042_red_width}}
\end{figure}

We observe that spiral arm width usually varies along the spiral arm (see Figure~\ref{fig:NGC1042_red_width}), which is not a new result~\cite{Savchenko2020}. In the mentioned work a linear function of radius was used to describe the width variation, despite that noticeable deviation from the linear function can be seen in their figure~7. However, in our results we mostly observe fairly good agreement with the linear behaviour. In some cases we see deviations from the linear behaviour, often in the form of temporary increase of width at some part of the arm, but it probably can be connected to the fact that some minor spurs which were not traced and masked properly, can pass near the spiral arm, temporary increasing its measured width.

We then compare which function fits observed spiral arms best. We can define width variations as a function of $r$ or as a function of $\psi$ as well, so we check both options. Specifically, we examine the linear function of $r$, linear function of $\psi$ and exponential function of $\psi$ (which is equal to the linear function of $r$ for the case of logarithmic spiral shape of spiral arms). We observe that linear function of radius fits width variations a little better than others (median $\chi^2 = 8.8$ versus 9.7 and 9.5 for linear and exponential function of azimuthal angle, respectively. Note that errors may be not normalized well, leading to absolute $\chi^2$ values less meaningful). In the following, we will ensure that there is no need to add modifications to this function, $w(r) = w_1 r + w_0$ (see Section~\ref{sec:2d_fitting}).

We calculate a parameter of zero-point contribution $w_\text{zpc}$ which characterises the relative importance of $w_0$. This parameter is calculated as $w_\text{zpc} = |w_0| / w(r_\text{avg})$, where $r_\text{avg}$ is an average between the radius at the beginning and at the end of the arm (for brevity, we also denote $w_\text{avg} = w(r_\text{avg})$). If zero-point is close to zero, then the width of spiral arm is mostly defined by $w_1 r$ term and is nearly proportional to $r$, then $w_\text{zpc}$ will be close to zero. Otherwise, $w_0$ cannot be ignored; for example, if $w(r)$ is constant, then $w_0$ contribution is dominant and $w_\text{zpc} = 1$.

Here we primarily consider only long spiral arms ($l_\psi \geq 180^\circ$) because data on width is noisy, and, for short arms and spurs, function parameters are determined poorly. Moreover, as these small features span across a small range of radii, some degree of degeneracy between $w_1$ and $w_0$ is inevitable.

Our results depend on the method of disc subtraction (see Section~\ref{sec:extraction}). Using $q = 0.1$ disc, we find that median $w_\text{zpc}$ is 19\%. By this reason, we consider $w_0$ parameter to be less important than $w_1$, and we see the median $w_1$ is 0.30, also being the same for grand-design subsample and for multi-armed. The median $w_\text{avg} / r_\text{avg}$ is 0.37, with the small difference between grand-design (0.36) and multi-armed galaxies (0.39). If we consider short spirals as well, then median $w_\text{zpc}$ increases to 35\%, and median $w_1$ becomes 0.28, albeit with a higher scatter. For spiral arms in ultraviolet, we get $w_1$ equal to 0.14 and $w_\text{zpc}$ of 39\%. The width in UV is much more noisy than in IR, due to the clumpiness of images.

However, results are somewhat different if we use images of spiral structure that were prepared using the decomposition disc. In this case, median $w_1$ becomes significantly lower, with the median value of 0.19, and $w_\text{zpc}$ is 34\% and the median $w_\text{avg} / r_\text{avg}$ is 0.29. As one can see in Figure~\ref{fig:profiles_examples}, spiral arm brightness is noticeably lower when decomposition disc is subtracted, which is the most likely reason for the said difference. At the ridge-line, the effect may be relatively small, but moving away from the ridge-line, spiral arm brightness will fall off more rapidly, and spiral arms appear narrower. 

In~\cite{Savchenko2020}, the median slope coefficient of $w(r)$ function was found to be 0.12. There are different reasons for such discrepancy: starting from purely technical ones, in their work perpendicular width is measured instead of radial one; for their average $\mu \approx 15^\circ$, perpendicular width is a factor of $\cos \mu \approx 0.97$ of the radial one. Secondly, they define width in a different way than we do: from the Equation 1 in~\cite{Savchenko2020}, their adopted width is essentially $2\sqrt{2}\sigma$ for symmetric Gaussian, whereas Gaussian FWHM used in our work is $2 \sqrt{2 \ln{2}} \sigma$, which is nearly 17\% smaller. Which is more important, however, it is that value of spiral arm width depends significantly on the estimation of underlying disc profile. Prior to analysing spiral structure, authors of~\cite{Savchenko2020} also estimate and subtract underlying disc profile, using a method that differs from both our methods (Section~\ref{sec:extraction}). Due to this reason, it is possible that their estimates of the disc brightness are systematically higher than in our work. As we discussed earlier in this Section, this may cause spiral arm brightness to fall off more rapidly when moving off the ridge-line of the spiral arm, and make spiral arms appear narrower. Finally, our work deals with the different wavelength range than~\cite{Savchenko2020}, whereas it is known that spiral arm width depends considerably on wavelength~\cite{Marchuk2024b, Chugunov2025}, and for stellar 3.6 $\mu$m images it is even higher than for original 3.6 $\mu$m.

Among all spirals, including short ones and spurs, we found that 88\% of them have width increasing to the periphery of a galaxy using $q = 0.1$ disc, and 82\% using the decomposition disc. This number is consistent with~\cite{Savchenko2020} where fraction 85.8\% is reported, and with the earlier result of~\cite{Honig2015} where all of 4 galaxies examined were found to have increasing widths of spiral arms to the periphery.

\subsection{Brightness profiles across spiral arm}
\label{sec:fine_profile}
Speaking about the brightness profile across the spiral arm, we will refer to the radial profile (along the line of constant $\psi$). Specifically, we consider its skewness (measure of how much the outer part of the arm is more or less extended then the inner), and how much the overall shape of the profile deviates from the Gaussian: does it have sharper or flatter peak. In statistics, the described properties of distributions can be related, essentially, to the 3rd and 4th central moments (skewness and kurtosis). Some of previously discussed parameters correspond to 1st moment (the ridge-line of spiral arm, essentially representing the center position of the slice $r$ at a given $\psi$ which can be treated as mean) and 2nd moment (width of the spiral arm, which can be related to variance). This consideration can justify that, previously, we ignored skewness and sharpness of the peak, as these parameters can be treated as ``fine tuning'' to the general spiral arm parameters, which can be approximated by simple Gaussian function.

An example of a function which can have different kurtosis is S{\'e}rsic function~\cite{Sersic1968}, which peak sharpness depends on its parameter $n$ and at $n = 0.5$ it becomes Gaussian. Moreover, it is often used in astronomy to describe surface brightness distributions~\cite{Graham2005}. It can be easily modified to have non-zero skewness: one can use asymmetric S{\'e}rsic function which would have different $r_e$ to the inner and to the outer side from the peak, and we will use this function to analyse the profile.

\begin{figure}[H]
\centering
\includegraphics[width=0.99\textwidth]{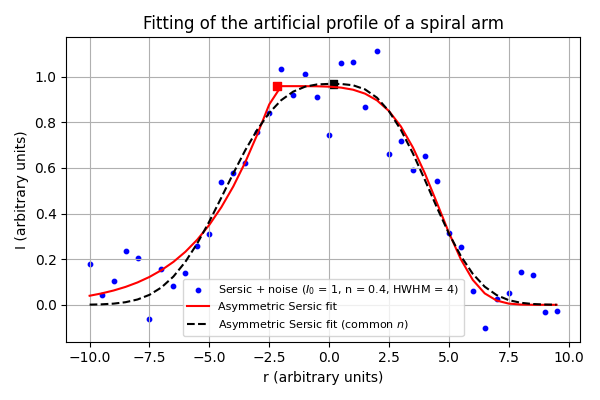}
\caption{This picture illustrates the downsides of using asymmetric S{\'e}rsic function with $r_e$ and $n$ both independent in inner and outer side of the arm. The fitted ``data'' (blue dots) is artificial, representing a slice of spiral arm and produced by symmetric S{\'e}rsic function with added random noise. It has peak brightness of 1, HWHM of 4 (both are in arbitrary units), has a peak at zero and $n = 0.4$. If one fits this data with asymmetric S{\'e}rsic function with both independent $r_e$ and $n$, it yields a function (red solid line) with different $n$ at both sides, strong skewness (outer half-width is more than 3 times exceeding inner half-width) and misplaced peak (marked by square; misplacement is more than 1/2 HWHM). Despite such a function fits observed data points well, parameters inferred from it are wrong. Similar kind of resulting model --- with high $n$ and low $r_e$ in one side and low $n$ and high $r_e$ in another, appeared in fit results for some images in the previous works, which is probably a sign of a similar problem. If one uses asymmetric S{\'e}rsic function with the same $n$ for inner and outer part for fitting, it yields parameters that are much more close to true.\label{fig:Asymm_Sersic}}
\end{figure}

In~\cite{Chugunov2024, Marchuk2024b, Chugunov2025}, however, we used a function which radial profile is asymmetric S{\'e}rsic function with not only $r_e$, but also $n$ defined independently for the inner and the outer side of arm. Now, we can show that using two independent $n$ in a function to fit real data can result in issues with the parameters of the fine profile. We illustrate this in Figure~\ref{fig:Asymm_Sersic}. In some cases, some of the measured parameters differ noticeably from the ones that were set originally, despite overall brightness distribution looking more or less similar to observed. However, this does not invalidate our conclusions in previous works. First, only the parameters of a fine profile (the skewness and S{\'e}rsic indices) are prone to instability, and no conclusions were based on measurements of these parameters. Secondly, in~\cite{Chugunov2024} the shape of spiral arm was fixed prior to decomposition, which prevents displacing the centre of a slice in a model. Meanwhile, in~\cite{Savchenko2020}, it is mentioned that asymmetric Gaussian function allows one to fit slices well without overfitting.

Therefore, in this step we fit slices with 1) an asymmetric Gaussian function and then 2) asymmetric S{\'e}rsic function with common $n$ in both sides. (Equation~\ref{eq:Asymm_Sersic}), gradually moving towards more complex functions.

\begin{linenomath}
\begin{equation}
	\label{eq:Asymm_Sersic}
	I_\bot (\rho) = 
    \begin{cases}
        I_0 \times \exp \left(-\ln (2) \times \left[\frac{\rho}{w_\text{in}}\right]^{1 / n}\right), & \rho < 0 \\
        I_0 \times \exp \left(-\ln (2) \times \left[\frac{\rho}{w_\text{out}}\right]^{1 / n}\right), & \rho \geq 0
    \end{cases}
\end{equation}
\end{linenomath}

\begin{figure}[H]
\centering
\includegraphics[width=0.99\textwidth]{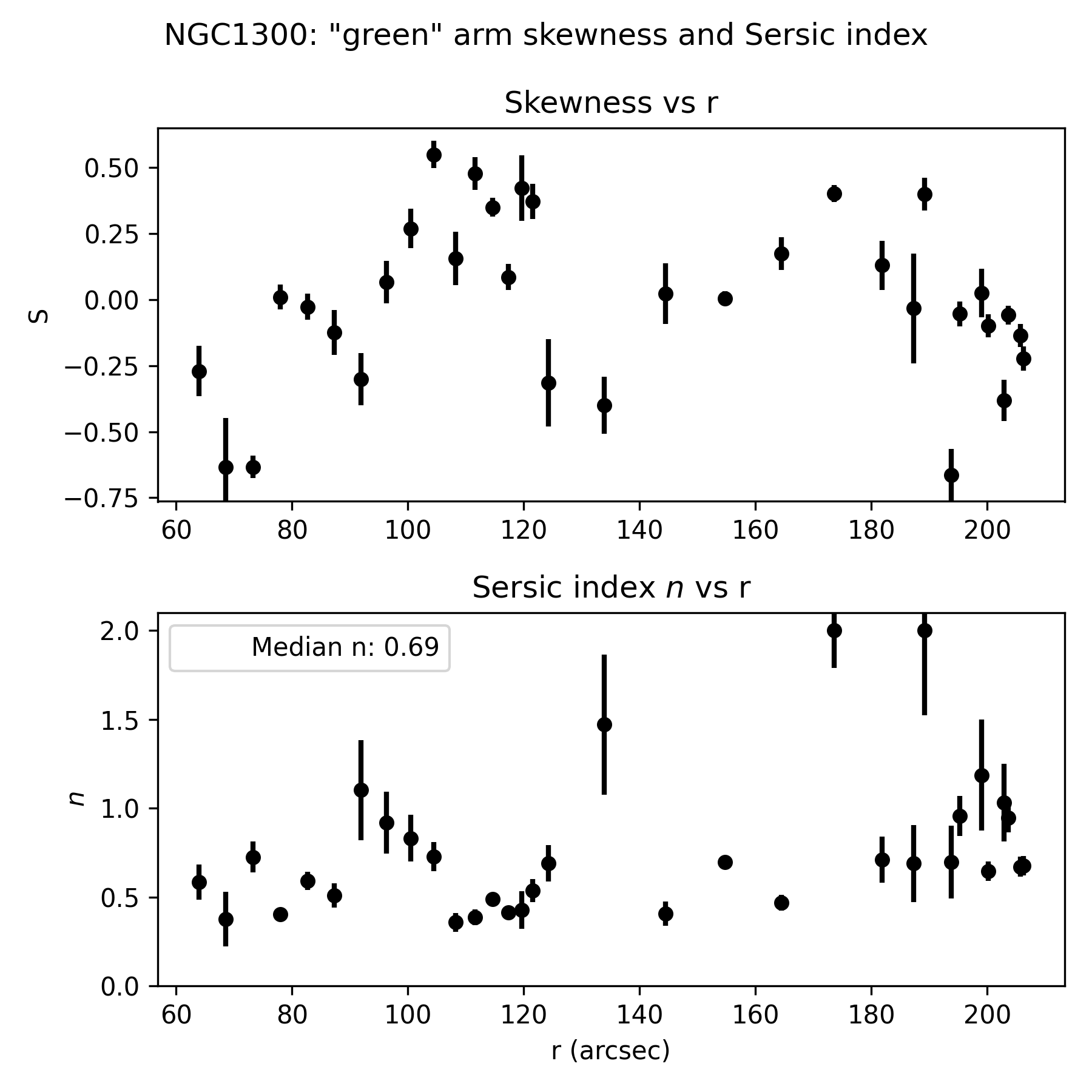}
\caption{An example of skewness $S$ and S{\'e}rsic index $n$ behavior in a spiral arm of NGC~1300, as a function of radius $r$.\label{fig:NGC1300_green_fine_profile}}
\end{figure}

Here, $\rho$ is the distance from the centre of the arm slice, $w_\text{in}$ and $w_\text{out}$ are inner and outer half-widths, respectively. Note that the S{\'e}rsic function presented here is not in its classical form, and $b_n$ coefficient is replaced with $\ln(2)$, and therefore $w_\text{in}$ and $w_\text{out}$ are not half-light radii, but HWHM (see Section 6.3.4 in~\cite{Marchuk2024b} for the reasoning to not use half-light radii). We remind that asymmetric Gaussian function is just a special case when $n = 0.5$. At this point, we are interested in obtaining profile skewness, defined as $S = \frac{w_\text{out} - w_\text{in}}{w_\text{out} + w_\text{in}}$, and S{\'e}rsic index $n$. In Figure~\ref{fig:NGC1300_green_fine_profile}, we show the typical behaviour of skewness $S$ and S{\'e}rsic index $n$.

\begin{figure}[H]
\centering
\includegraphics[width=0.99\textwidth]{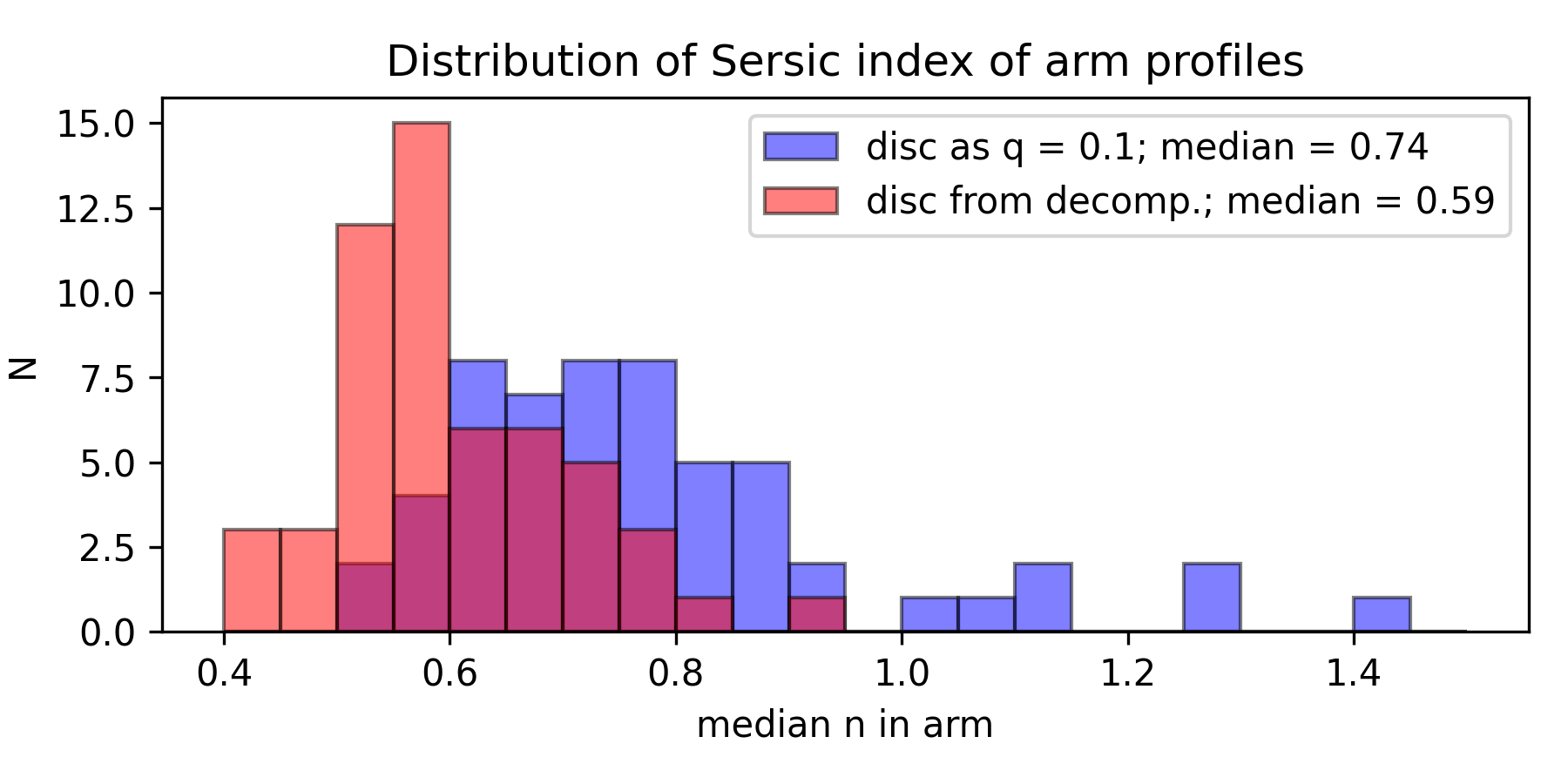}
\caption{Distribution of median S{\'e}rsic indices of individual arms. Blue columns show results for spiral arms extracted with $q = 0.1$ disc, red columns --- with decomposition disc.\label{fig:Sersic_index_distribution}}
\end{figure}

However, the only consistent result that we observe is that the scatter in measured parameters is high. The noise is significant, and therefore parameters that describe such a fine structure cannot be accurately derived from individual slices. We can only suspect that, in some cases, a large-scale trend of skewness variation along the arm exists (which can be connected with the location of the corotation radius; see Section~\ref{sec:width_gradients}). Concerning the S{\'e}rsic index, the median values for spiral arms is usually close to 0.5 or higher, and in most cases lower than 1 (see Figure~\ref{fig:Sersic_index_distribution}), but the scatter between individual slice measurements is high. Moreover, the values is not exactly the same when different methods of disc subtraction are used (Section~\ref{sec:extraction}); obviously, the small difference in subtracted disc profiles affect the periphery of the spiral arm relatively strongly, which is a defining feature for $n$. At this point, we abstain from making any decisive conclusions. We will have an opportunity to test this conjecture by fitting an entire spiral arm with 2D functions (Section~\ref{sec:2d_fitting}).

\section{Constructing a photometric model of a spiral arm}
\label{sec:photometric_model_preliminary}

Summarising all results above, we can construct two-dimensional photometric model of a spiral arm. In previous works~\cite{Chugunov2024,Marchuk2024b,Chugunov2025}, we used our own model for decomposition and it worked well for different samples of images. However, it was constructed rather arbitrarily, and some possible room for improvement is now evident. When constructing a function, we emphasize that parameters should be, as much as possible, easy to interpret and to control. It also would be useful to have some simple function variant, appropriate for fitting spurs and for cases when the resolution is poor; therefore, one has to know which parameters' effect on the function is minimal, so they can be omitted without losing too much in fit quality.

Now we will describe some tentative model which properties are in accordance with properties of spiral arms that we measured before. Then, we will perform some checks and validations to confirm or modify some of function features. Here, we will describe our function briefly and qualitatively; the full and more formal description will be presented in Section~\ref{sec:photometric_model_final} for a finally established function.

Generally, the overall design of the function follows our previous works starting from~\cite{Chugunov2024}: it should define each spiral arm individually, modelling the light distribution of the arm in the galactic plane, using polar coordinates. It is convenient to express it with shape function $r(\psi)$ that defines the ridge-line of the spiral arm, the function $I_\parallel$ which defines the light profile along the ridge-line, and $I_\bot$ defining how brightness changes across the spiral arm. More precisely, $r(\psi)$ is a polynomial-logarithmic function, $I_\parallel$ is exponential function of $r$ modified by a flat-top window function of $\psi$, yielding a smooth transition to zero near the ends, and $I_\bot$ is an asymmetric S{\'e}rsic function which width and skewness change linearly with radius.

For the following checks, we have to select a preliminary 2D function, relying on our previous results. A function that we started from, is almost similar to the function that we finally adopt (Section~\ref{sec:photometric_model_final}) following our validation described in Section~\ref{sec:2d_fitting}. The only difference is the part of $I_\parallel$: this function represented as a product of two functions $I_\parallel(r(\psi), \psi) = I_{\psi \parallel}(\psi) \times I_{r \parallel}(r)$, and $I_{\psi \parallel}(\psi)$ in our function for validation described in a following way:

\begin{equation}
    \label{eq:growth_cutoff_preliminary}
	I_{\psi \parallel}(\psi) =
	\left\{\begin{array}{ll}
		G(1, \psi_\text{start}, \psi_\text{growth}) & \psi < \psi_\text{start}\\
		1 & \psi_\text{start} \leq \psi < \psi_\text{end}\\
		G(1, \psi_\text{end}, \psi_\text{cutoff}) & \psi \geq \psi_\text{end}
	\end{array}\right.\
\end{equation}

Here, $G(a, b, c)$ is a Gaussian function with peak value $a$, peak location $b$ and standard deviation $c$.

\subsection{Validation by 2D fitting}
\label{sec:2d_fitting}
To check the goodness of our models and the appropriateness of the set of parameters, we perform the 2D fitting of straightened spiral arms with our models. The shape of spiral arms' ridge-line in this case is kept out of discussion.

\begin{figure}[H]
\centering
\includegraphics[width=0.99\textwidth]{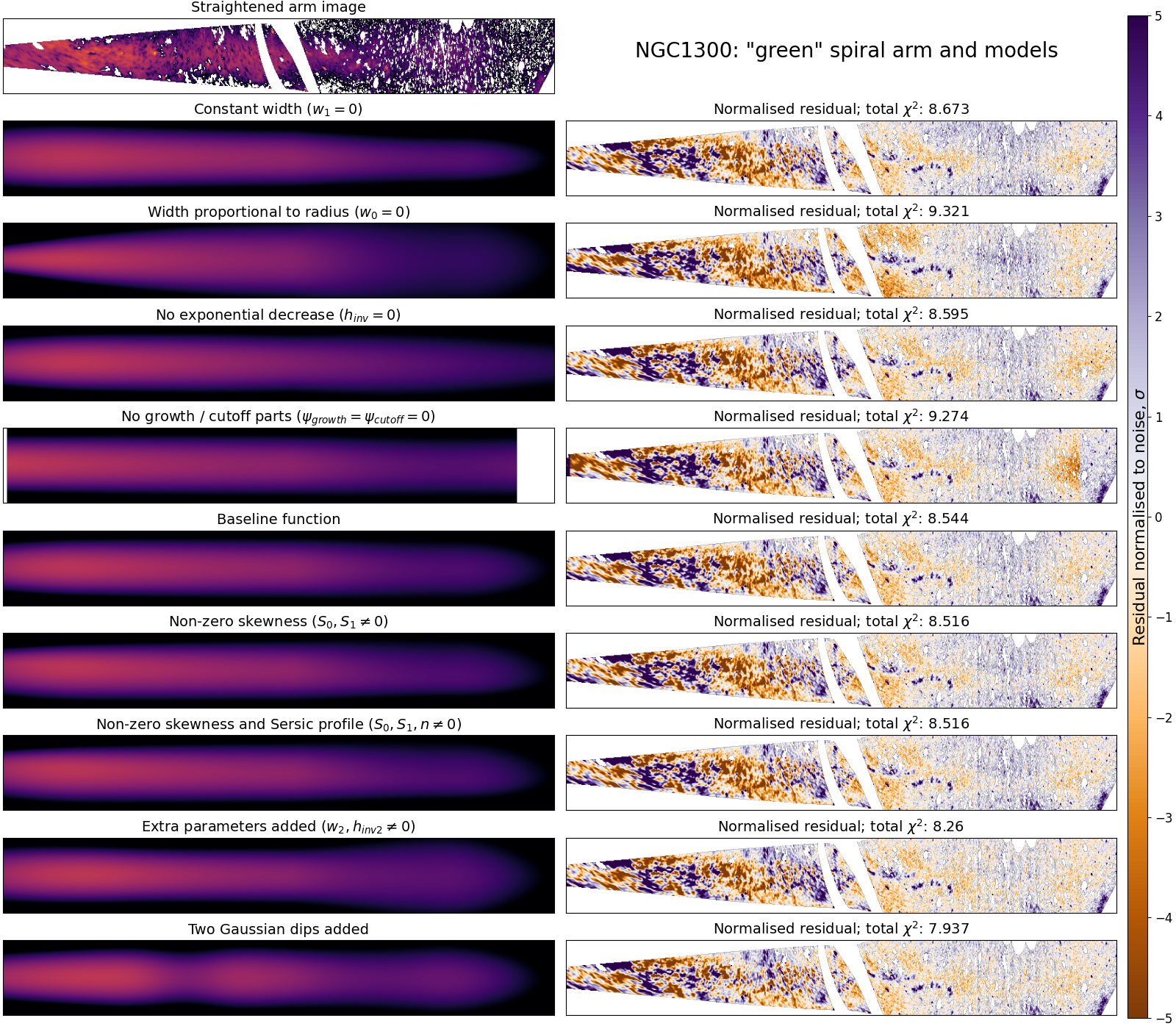}
\caption{An example of a straightened spiral arm of NGC~1300, fitted by different modifications of a baseline function. The image of the arm is located at the top, the sequence of a models is in the left column, residuals normalised to noise map are on the right.\label{fig:NGC1300_green_multiple_fits}}
\end{figure}

More specifically, we fit our adopted function in a few variants and examine how strongly our modifications affect $\chi^2$ of the fit. As a baseline function, we take our adopted function (Section~\ref{sec:photometric_model_preliminary}) with a constant zero skewness and with Gaussian profile, and consider modifications of this variant. The list of modifications that were fitted is the following, and in Figure~\ref{fig:NGC1300_green_multiple_fits} there is an example of different functions fitted to a single straightened arm.

In particular, we test adding higher order polynomial coefficients to $w_r(r)$ and $I(r)$ function which we do not intend to add into our final function. Instead, we do this to ensure that the addition of these parameters does not significantly improve fit quality, and they can be safely excluded from the model. With the addition of these parameters, Equations~\ref{eq:I_r_parallel} and~\ref{eq:w_loc} have the following form (Equation~\ref{eq:w_I_add}):

\begin{equation}
    \label{eq:w_I_add}
	\begin{array}{ll}
		I_{r\parallel}(r) = I_0^\text{sp} \times e^{-(h_\text{inv} r + h_\text{inv2}r^2)}\\
		w_\text{loc} = w_2 r(\psi)^2 + w_1 r(\psi) + w_0
	\end{array}
\end{equation}

\begin{itemize}
    \label{list:2d_funcs}
    \item (1) Function with constant $I_\parallel$ except growth/cutoff parts (effectively, $h_\text{inv} = 0$)
    \item (2) Function with sharp transition of $I_\parallel$ ($\psi_\text{growth} = \psi_\text{cutoff} = 0$)
    \item (3) Function with constant width ($w_1 = 0$)
    \item (4) Function width proportional to radius ($w_0 = 0$)
    \item (5) Function with linearly changing skewness (nonzero $S_0, S_1$)
    \item (6) Function with linearly changing skewness and S{\'e}rsic profile (nonzero $S_0, S_1, n$)
    \item (7) Function with quadratic $w_\text{loc}(r)$ and $I_\parallel$ (additional parameters $w_2$ and $h_\text{inv2}$, see Equation~\ref{eq:w_I_add})
    \item (8) Optional, if dips are present: function with 1 or 2 Gaussian brightness dips (Equation~\ref{eq:dips})
\end{itemize}

With this set of functions, we can measure the importance of a number of individual parameters, from the informaton on how fit quality changes when function without that parameter is being used (see~\ref{tab:chisq_comp}).

\begin{table}[H]
\caption{Mean ratios of $\chi^2$ between fits with different functions for the entire sample of spirals.\label{tab:chisq_comp}}
		\begin{tabularx}{\textwidth}{CCCCCCCCCC}
			\toprule
            & Baseline function & (1) & (2) & (3) & (4) & (5) & (6) & (7) & (8) \\
            Add. par. & $10$ & $-1$ & $-2$ & $-1$ & $-1$ & $+2$ & $+3$ & $+2$ & $+3$/$+6$ \\
            \midrule
            \multicolumn{10}{c}{$\langle\chi^2 / \chi^2_\text{b}\rangle$, $q = 0.1$ disc} \\
            \midrule
            All & 1 & 1.017 & 1.196 & 1.120 & 1.018 & 0.991 & 0.983 & 0.972 & 0.817 \\
            $l_\psi~>~180^\circ$ & 1 & 1.012 & 1.195 & 1.086 & 1.021 & 0.990 & 0.985 & 0.961 & 0.817 \\
            \midrule
            \multicolumn{10}{c}{$\langle\chi^2 / \chi^2_\text{b}\rangle$, decomposition disc} \\
            \midrule
            All & 1 & 1.021 & 1.137 & 1.043 & 1.008 & 0.987 & 0.982 & 0.960 & 0.829 \\
            $l_\psi~>~180^\circ$ & 1 & 1.034 & 1.118 & 1.024 & 1.010 & 0.987 & 0.979 & 0.959 & 0.829 \\
            \bottomrule
		\end{tabularx}
	\noindent{Numbers (1)--(8) denote the functions described in text above (Section~\ref{list:2d_funcs}). ``Add. par.'' is a number of additional parameters compared to baseline function. $\langle\chi^2 / \chi^2_\text{b}\rangle$ is the average ratio between $\chi^2$ for a given function and $\chi^2_\text{b}$ for a baseline function; $l_\psi > 180^\circ$ refers to the same parameter, but for long spiral arms.}
\end{table}

We check results for both our methods of extracting spiral arms, and present values for all sample of spirals, as well as for only long spiral arms. In any case, the results do not differ much. In particular, we found that the addition of brightness dip (column 8), where appropriate, offers the most significant increase of $\chi^2$ among all modifications. While it makes a function more complex by 3 parameters per dip, adding relevant number of dips (1 or 2) decreases $\chi^2$ by 17--18\%, on average (nearly 5\% per parameter, considering the number of arms with 1 and 2 dips). On the other hand, adding $h_\text{inv2}$ and $w_2$ (column 7) into a model makes fit improvement much more modest (3--4\% by 2 parameters), albeit it can be moderately useful in general, adding some fine-tuning to the profile. Nevertheless, $h_\text{inv2}$ and $w_2$ themselves have dubious physical meaning and are harder to control, and we prefer including the option to add brightness dips in a model.

Addition or the skewness coefficients and $n$ (column 6) decreases $\chi^2$ by only 2--3\% combined. Despite their small contribution to the fit quality, their addition does not lead to the overfitting, as we made sure comparing BIC statistics. The reason to still include these parameters into our model is that they potentially have physical meaning: if the skewness of spiral arm profile changes with radius, it can be helpful to detect the location of corotation radius (see~\cite{Marchuk2024a} and Section~\ref{sec:width_gradients}). We recommend not using this parameters by default, but add them purposefully.

In the case of setting $I_\parallel$ constant except the growth and cutoff region (column 1), removing this parameter (effectively setting $h_\text{inv} = 0$) has a surprisingly small effect, increasing $\chi^2$ by only 1--3\%. In the same time, setting edges of spiral arm sharp, i.e. $\psi_\text{growth} = \psi_\text{cutoff} = 0$ (column 2), makes fit results significantly worse, increasing $\chi^2$ by more than 10\% for just 2 parameters. All these parameters are related to $I_\parallel$, and this result can be interpreted as another sign of that exponential decrease does little to properly describe the profile of spiral arm. Dealing with spurs, which are expected to span a small range of radii, it is probably the most effective to set $h_\text{inv} = 0$ for them.

Concerning the behaviour of width of spiral arms, there is a similar phenomenon of that parameters do not have the same importance. Making spiral arms of constant width (column 3), effectively setting $w_1 = 0$, leads to a noticeable decrease of a fit quality, with $\chi^2$ increasing by a few percent; in the same time, making spiral arm width strictly proportional to the radius, thus setting $w_0 = 0$ (column 4) increases $\chi^2$ only by 1--2\%. This aligns with our previous observation on that the linear increase of the width is definitive to the spiral arm width anywhere at the arm (see Section~\ref{sec:widths}). Therefore, one can use $w_0 = 0$ for short spurs, which can also prevent a possible degeneracy between $w_0$ and $w_1$ (again, see Section~\ref{sec:widths}).

As we were unable to derive properties of the perpendicular profile of spiral arms by slice fitting (Section~\ref{sec:fine_profile}), we use the results of 2D fits to measure them. This way, we obtain S{\'e}rsic indices $n$ of profiles along the long spiral arms ($l_\psi > 180^\circ$) comparable with slice fitting, with median values somewhat depending on disc subtraction method (Section~\ref{sec:extraction}): for $q = 0.1$ disc, the median $n = 0.67$, and for decomposition disc, the median $n = 0.59$. In any case, for most arms $0.4 < n < 0.9$. The overall distribution is looking similar to that obtained with slice fitting (Figure~\ref{fig:Sersic_index_distribution}), so the conclusion is that spiral arms profiles have more or comparably pronounced peak than the Gaussian function, but less pronounced than the exponential. Concerning the skewness distribution, both average value along the arm and trend, our measurements show a large scatter around zero value, depending slightly on disc subtraction method.

\subsubsection{Selecting a growth and cutoff function}
\label{sec:growth_and_cutoff}
In a similar way, we examine which function describes growth and cutoff parts of the spiral arms the best. Here we introduce some formalism connected to $I_\parallel$: we define it as a function of $\psi$ and $r(\psi)$ and expand it as $I_\parallel(r(\psi), \psi) = I_{r\parallel}(r(\psi)) \times I_{\psi\parallel}(\psi)$. Here, $I_{r\parallel}$ is the part defining exponential decrease with radius, whereas $I_{\psi\parallel}$ is a truncation function in a form of a flat-top window function. Along some part of the spiral arm, where exponential decrease is in place, it equals 1, and it provides a smooth transition from zero at the beginning of the arm (growth) and to zero at the end of the arm (cutoff). The general form of this function is provided in~\ref{eq:growth_cutoff_general}.

\begin{equation}
    \label{eq:growth_cutoff_general}
	I_{\psi \parallel}(\psi) =
	\left\{\begin{array}{ll}
        0 & \psi < 0\\
		T(\psi) & 0 \leq \psi < \psi_\text{growth}\\
		1 & \psi_\text{growth} \leq \psi < \psi_\text{cutoff}\\
		T'(\psi) & \psi_\text{cutoff} \leq \psi \leq \psi_\text{end}\\
        0 & \psi > \psi_\text{end}
	\end{array}\right.\
\end{equation}

Values $\psi_\text{growth}, \psi_\text{cutoff}, \psi_\text{end}$ are function parameters that define azimuthal angles where growth part ends, cutoff part begins and the entire arm ends, respectively. At this step, our aim is to select a transition function $T$ which defines how $I_{\psi\parallel}$ exactly goes from 0 to 1 (and $T'$ from 1 to 0). We test 4 different functions: linear, quadratic, Gaussian (which cannot reach zero, but approaches close enough for any practical purpose) and a cubic polynomial $3(\psi / \psi_\text{growth})^2 - 2(\psi / \psi_\text{growth})^3$ which derivative at 0 and $\psi_\text{growth}$ is both zero.

We fit our baseline model with each of these functions used for growth and cutoff parts, and compare $\chi^2$. Note that, in all these cases, there is only two parameters needed, one defining the length of growth part, and one for cutoff. We found that, with all truncation functions, model can fit observed light distribution equally good, with average $\chi^2$ being within $\pm$1\% for all functions.

\subsection{Proposed photometric function}
\label{sec:photometric_model_final}
Our model defines the 2D distribution of light $I(r, \psi)$ as a function of polar galactocentric coordinates in the galactic plane. Therefore, among the parameters needed to define the 2D light distribution in the spiral arm, there are coordinates of the galaxy centre $[X_0, Y_0]$ and the galactic plane orientation characterised by the position angle $\text{PA}$ and ellipticity $\text{ell}$.

Each arm has a shape function $r(\psi)$, which defines the ridge-line of a spiral arm; we remind that, for a given $\psi$, this function defines $r$ where brightness is highest, and the ridge-line of the arm is a curve $[r(\psi), \psi]$ in polar coordinates. This is an additional parameter of $\varphi_0$ which is an azimuthal angle in the galactic plane, where spiral arm starts, serving as an origin of $\psi$.

When the shape function is defined, our distribution of light $I(r, \psi)$ can be represented as a product of two more simple functions: $I_\text{sp}(r, \psi) = I_\parallel(r(\psi), \psi) \times I_\bot(r - r(\psi), r(\psi))$. Here, $I_\parallel$ is a term describing the light distribution along the spiral arm ($\psi$ and $r(\psi)$ are not independent, but it is more convenient to use both parameters as variables). $I_\bot$ describes the brightness profile across the spiral arm, in a radial direction: $r - r(\psi)$ is a radial distance from the point to the ridge-line of the arm (hereafter, $\rho = r - r(\psi)$), and the presence of a second parameter $r(\psi)$ means that brightness profile across the arm depends on radius.

Next, we provide definitions for the components of our model. Based on the results in Section~\ref{sec:shapes}, the shape function of spiral arm $r(\psi)$ is, by default, defined as a polynomial-logarithmic spiral (Equation~\ref{eq:r_psi}) with $N = 2, 3, 4$ depending on spiral arm length ($l_\psi < 90^\circ$, $90 \leq l_\psi < 180^\circ$, $l_\psi > 180^\circ$ respectively). If one suspects the presence of bends, the piecewise modification of this function should be used instead (Equation~\ref{eq:r_psi_bends}) with 1 or 2 bends and $N = 2$ or 3 at each smooth part. Sometimes, bends can be seen directly on images, but if not, they are often connected with spiral arm bifurcations, or passages near the ends of the bar.

For $I_\parallel$, we present this function as a product of two functions, for convenience (see also Section~\ref{sec:growth_and_cutoff}): $I_\parallel(r(\psi), \psi) = I_{r\parallel}(r(\psi)) \times I_{\psi\parallel}(\psi)$. $I_{r\parallel}(r(\psi))$ defines the exponential decrease (or, in some cases, increase) with radius (Equation~\ref{eq:I_r_parallel}):

\begin{linenomath}
\begin{equation}
	\label{eq:I_r_parallel}
    I_{r\parallel}(r) = I_0^\text{sp} \times e^{-h_\text{inv} r}\,.
\end{equation}
\end{linenomath}

This is much similar to the exponential disc profile, but instead of $r/h$ under the exponent, we prefer to use $h_\text{inv} r$, so $h_\text{inv} = 1 / h$, and function does not change abruptly when $h_\text{inv}$ is varied near zero value. As concluded in Section~\ref{sec:2d_fitting}, one can use a constant $I_{r\parallel}(r)$ in this form without a significant lose in fit quality, and therefore one can safely assume $h_\text{inv} = 0$ for spurs.

On the $I_{\psi\parallel}$, we have performed special tests in Section~\ref{sec:growth_and_cutoff}. Despite we found that Gaussian growth and cutoff allows one to obtain the best $\chi^2$, its advantage is very small compared to other functions. However, this function has a drawback of not reaching zero at any finite distance, and then it is impossible to characterise where spiral arm ends, if modelled by this function. Therefore, we select a cubic polynomial that reaches zero at a finite distance, which was arbitrarily chosen in~\cite{Chugunov2025}. If we compare models where this function was used for growth and cutoff, and models where Gaussian was used, the difference between mean $\chi^2$ for them is just 0.5\%. Thus, we adopt a function for $I_{\psi\parallel}$ as from Equation~\ref{eq:growth_cutoff_general}, with $T$ and $T'$ defined as following (Equation~\ref{eq:I_psi_T}):

\begin{equation}
    \label{eq:I_psi_T}
		\left\{\begin{array}{ll}
		T(\psi) = 3 \left(\frac{\psi}{\psi_\text{growth}}\right)^2 - 2 \left(\frac{\psi}{\psi_\text{growth}}\right)^3\\
        T'(\psi) = 3 \left(\frac{\psi_\text{end} - \psi}{\psi_\text{end} - \psi_\text{growth}}\right)^2 - 2 \left(\frac{\psi_\text{end} - \psi}{\psi_\text{end} - \psi_\text{growth}}\right)^3
	\end{array}\right.\
\end{equation}

This definition of $T$ and $T'$ makes $I_{\psi\parallel}$ a function with a continuous first derivative which yields smooth 2D distribution of light, and a compact support which makes it easier to control and interpret.

However, the function in this form does not account for possible dips in brightness. For these cases, we add a multiplier $D(\psi)$ whenever needed, so formally we have $I_\parallel$ defined by Equation~\ref{eq:dips}:

\begin{equation}
    \label{eq:dips}
		\begin{array}{ll}
		I_\parallel(r(\psi), \psi) = I_{r\parallel}(r(\psi)) \times I_{\psi\parallel}(\psi) \times D(\psi)\\
        D = 1 / [1 + \sum_i G(A_i, \psi_i^c, \psi_i^w)]
	\end{array}
\end{equation}

Here, $G(a, b, c)$ is a Gaussian function with peak value $a$, peak location $b$ and standard deviation $c$. $\sum_i$ means that, in principle, it is possible to add as much Gaussian dips as needed, but we observe arms with only 1 or 2 dips present. It is easy to decide, if this modification is needed in each special case, thanks to the fact that dips are visible. If so, it is also easy to make an initial guess of their parameters.

As for $I_\bot$, we first define the local width $w_\text{loc}$ and the local skewness $S_\text{loc}$, both depending on $r(\psi)$. The dependence is linear for both parameters: $w_\text{loc} = w_1 r(\psi) + w_0$. $w_\text{loc}$ and $S_\text{loc} = S_1 r(\psi) + S_0$; $w_0, w_1, S_0, S_1$ are all function parameters.

Local width is an overall FWHM of the profile in radial direction at some point of the arm. Local skewness defines how HWHM in outer direction $w_\text{loc}^\text{out}$ relates to the HWHM in inner direction $w_\text{loc}^\text{in}$ from the ridge-line, which is presented in~\ref{eq:w_loc}:

\begin{linenomath}
\begin{equation}
	\label{eq:w_loc}
    \begin{cases}
        w_\text{loc} = w_1 r(\psi) + w_0\\
        w_\text{loc}^\text{in} = w_\text{loc} \frac{1 - S_\text{loc}}{2}\\
        w_\text{loc}^\text{out} = w_\text{loc} \frac{1 + S_\text{loc}}{2}
    \end{cases}
\end{equation}
\end{linenomath}

Then, $I_\bot$ itself is defined as following (Equation~\ref{eq:I_bot}):

\begin{linenomath}
\begin{equation}
	\label{eq:I_bot}
	I_\bot(\rho, r) = 
    \begin{cases}
        \exp \left(-\ln (2) \times \left[\frac{\rho}{w_\text{loc}^\text{in}}\right]^{1 / n}\right), & \rho < 0 \\
        \exp \left(-\ln (2) \times \left[\frac{\rho}{w_\text{loc}^\text{out}}\right]^{1 / n}\right), & \rho \geq 0
    \end{cases}
\end{equation}
\end{linenomath}

\begin{figure}[H]
\centering
\includegraphics[width=0.99\textwidth]{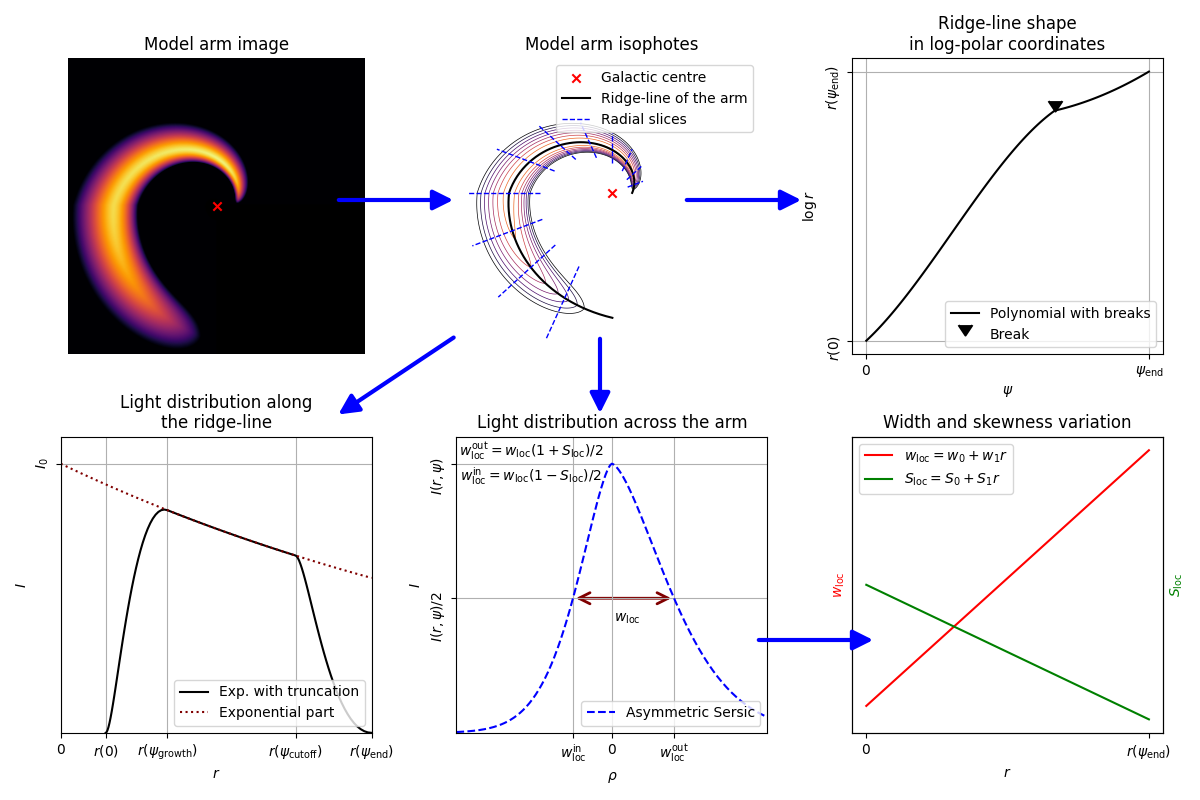}
\caption{This picture illustrates the properties of our 2D model. A possible implementation of our 2D model is shown in the top left, and its isophotes are in the top middle. Ridge-line of the spiral arm is superimposed on the isophotes image, and radial slices across the arm are also drawn. Top right shows the ridge-line in log-polar coordinates, defining the overall shape of the spiral arm. Lower left shows the light distribution along the ridge-line. Lower middle shows an example of radial profile across the spiral arm. This profile is particularly defined by the local width and local skewness, which variation is presented in the bottom right.\label{fig:arm_model_structure}}
\end{figure}

In Fig.~\ref{fig:arm_model_structure}, the properties of the model are summarised and the breakdown to simple functions is presented.

Finally, the resulting model, now justified, turns out to resemble our old models by most of the properties~\cite{Chugunov2024, Marchuk2024b, Chugunov2025}. However, there are some differences in the functional form of our new model. First, spiral arms in our new model can have bendings, i.e. abrupt changes of their pitch angle which can be seen as the generalization of our old function. Next, we define the width and skewness of the spiral arm as a function of radius, not azimuthal angle which is more justified physically and yields better fit results. The growth and cutoff function is also slightly different compared to previous works, and our new function also include a possibility to produce surface brightness dips in spiral arms. Finally, we identified and removed a redundancy of that two different S{\'e}rsic indices are used for two sides of the perpendicular spiral arm profile. Which is no less important, it is that we managed to identify the more and the less important parameters of a model. In turn, this allows us to adjust the function case-by-case: for example, when fitting spurs, dealing with bad resolution or being in no need of highly precise models, we can drop some of the least important parameters (technically, fixing some of them to zero or other default value). In particular, we determine how complex shape function can be, depending on the azimuthal length of spiral arms. Therefore, the actual number of parameters can differ case-by-case, but, for ordinary spiral arm without bendings and surface brightness dips, there can be 20 parameters total, comparable with the number in our previous works. Note that 4 of these parameters define the galactic plane and are shared with disc parameters, so there are 16 parameters unique for each spiral arm. For spurs, the number of unique parameters can be even limited to 9, which is not far from the very simple spiral arm model from~\cite{Lingard2020}. Almost all possible parameters of the spiral arm in our model have their own physical or geometrical meaning, and therefore model can be readily used for decomposition.

\section{Connection with the nature of spiral arms}
In the density wave theory framework, spiral galaxies are expected to have a corotation radius (CR), where the velocity of the spiral pattern is equal to the velocity of the disc~\cite{Dobbs2014}. The observed location and the possible multiplicity (see \cite{Marchuk2024c} and references therein) of the CR is one of the most important known indicators of the nature of spiral structure. It is known to be connected with various properties of spiral arms, including some of these which were measured in our work.

\subsection{Observable features}
\label{sec:observable_features}
Knowing the existence and locations of various observed features discussed below, we can compare them with known values of corotation radii in galaxies, collected from the literature in~\cite{Kostiuk2024}.

\subsubsection{IR -- UV offsets}
\label{sec:offsets}
If a density wave is present, then inside the CR, the velocity of matter is higher than of the spiral pattern, and the opposite is true outside CR. Then, one can consider a cloud of gas: when it approaches a spiral arm, star formation is triggered and extremely young stellar population can be observed. With time, this cloud moves relative to the spiral structure and its stellar population ages. It means that an azimuthal age gradient across the spiral arm is expected to take place, with a different directions inside and outside the CR. Locating this change of sign can be used for the determination of the CR, and different tracers of younger and older stellar population can be used for this purpose. For example, in~\cite{Tamburro2008} H I emission and 24$\mu$m radiation were used, corresponding to the offset between gas concentration and star formation. In~\cite{Yu2018b}, the difference of pitch angles between BVRI bands and 3.6 $\mu$m was found, caused by the said offsets. \cite{Miller2019} is another more multiwavelength study with a similar result.

In our data, we can use an offset between IR and UV images, with the former tracing old stellar population, and the latter showing the youngest. We employ the images of straightened spiral arms, prepared for both bands in the same coordinates, which allows us to measure offsets slice-by-slice, similar to the recent study~\cite{Kalita2025a}. Another possible option is to model spiral structure (possibly with decomposition) in both bands; in this case, offsets can be derived directly from model parameters. This approach is used for other galaxies in~\cite{Kostiuk2025}.

\subsubsection{Width gradients}
\label{sec:width_gradients}
A method of CR determination which uses a single-band photometric data was demonstrated in~\cite{Marchuk2024a}. The reasoning underlying this method is similar to the previous one (Section~\ref{sec:offsets}), and the key effect to be observed is that azimuthal brightness profile of the spiral arm is skewed in different direction inside and outside the CR. Just as for offset method, one can possibly utilize either slicing (as it was done in original work), or decomposition, if a model allows one to construct spiral arms with varying skewness.

\subsubsection{Bendings}
\label{sec:bendings}
Bendings in spiral arms are conspicuous features which can be examined on the possible connection with the resonances. A possible line of reasoning is that, effectively, CRs divide disc into two not fully connected parts. In particular, this can manifest itself in radial metallicity breaks, see~\cite{Scarano2013}), and it is not unlikely that abrupt changes in spiral arm parameters, including pitch angles, may be consistent with CR. Indeed, there is an example of M~51, which spiral arm bendings are coincident with CR positions\cite{Font2024}.

\subsubsection{Brightness dips in arms}
In~\cite{Pan2023} it is noted that, at CR, gas does not pass through the spiral arm. Therefore, one should expect that star formation rate is strangled at this radius, leading to the appearance of brightness dip. In the mentioned work, it is argued that these features are absent, and it is claimed to contradict the density wave theory. Although this reasoning offers a good prediction, the final conclusion of~\cite{Pan2023} is rather bold, given that only a single galaxy was examined and that one has a rather unusual spiral pattern.

Meanwhile, we often observe dips in spiral arms' surface brightness profiles, but they do not necessarily mark locations of CRs. Moreover, predicted dips connected with the suppression of star formation, are less likely to be observed at 3.6$\mu$m band, as this wavelength is dominated by old stars radiation. On the other hand, FUV images which represent the star formation distribution, are rather clumpy and the mentioned brightness dip is harder to be recognized.

\subsection{An example of NGC~4535}
\begin{figure}[H]
\centering
\includegraphics[width=0.99\textwidth]{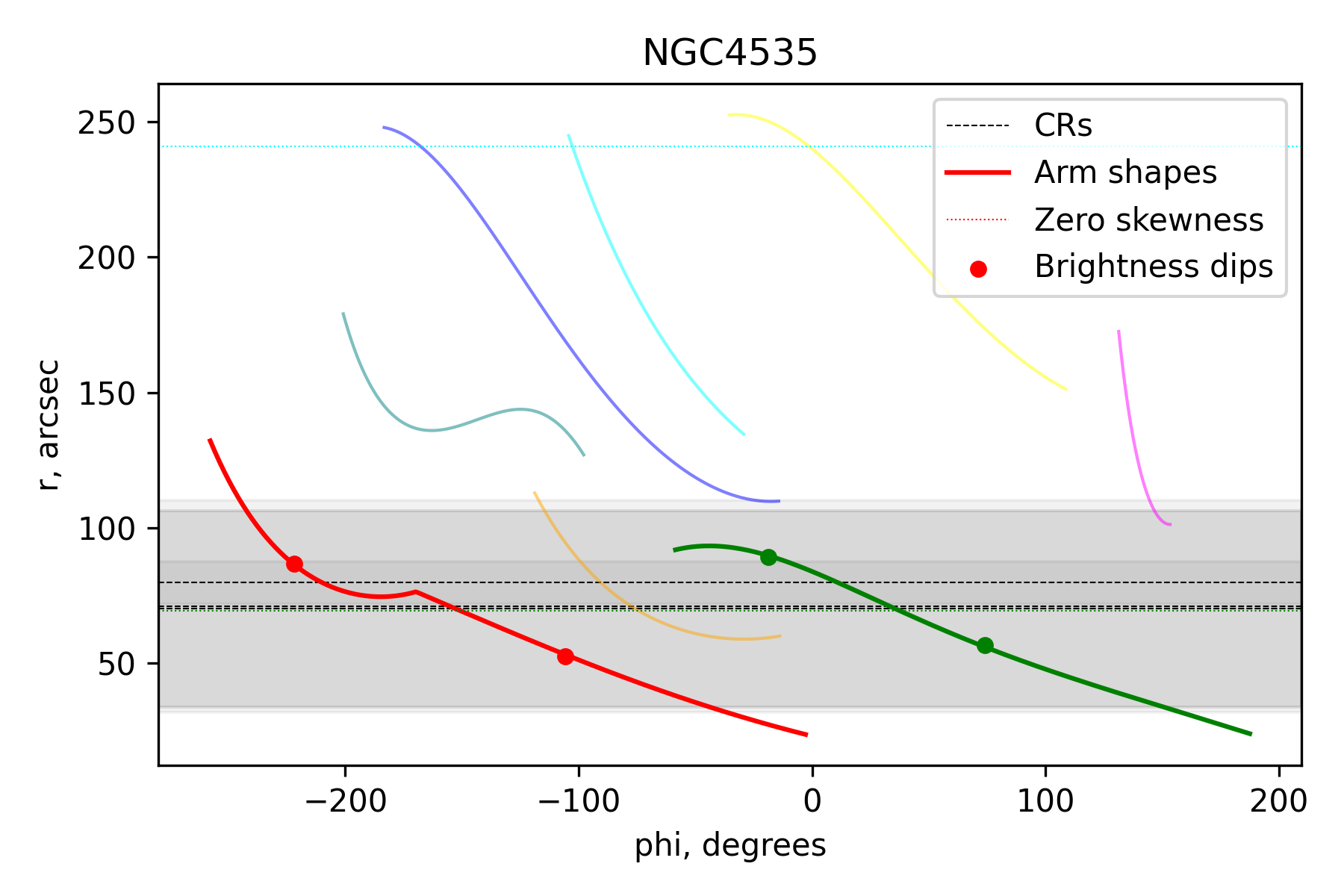}
\caption{This diagram shows the ridge-lines of spiral arms in $r$ -- $\psi$ coordinates for NGC~4535 as coloured curves, each one corresponding to a single spiral arm. Some of spiral arm features are also shown: specifically, the locations of brightness dips, when exist, are marked by points, and the located radii where skewness reach zero are shown as horizontal dotted lines of the corresponding colour. CR measurements from~\cite{Kostiuk2024} are also shown as black dashed lines, with their uncertainties presented with shaded areas.\label{fig:NGC4535_features}}
\end{figure}

Although each of all 19 galaxies has at least two different CR measurements, according to~\cite{Kostiuk2024}, NGC~4535 is an only object where all measurements (\cite{Elmegreen1995,Williams2021}) are consistent to at least some extent. In this galaxy, CR is estimated to be located 70--80 arcsec from the centre (however, some measurements are highly uncertain). This can be interpreted as a sign of that NGC~4535 possess a density wave in its disc. Also, there is another evidence for the CR being located at this radius. The key point of the Font-Beckman method~\cite{Font2011} is that the radial gas velocity is expected to reach zero at the resonance locations, which can be observed as zeros in residual velocity maps. In fact, residual velocity maps of NGC~4535 exist for HI and CO data (see Figures 5 and 6 in~\cite{Laudage2024}), and at 70--80 arcsec from centre, they are indeed close to zero. Therefore, this galaxy is a natural choice to test the connection of these features with a corotation resonance. In Figure~\ref{fig:NGC4535_features}, we present the ridge-lines of spiral arms as well as locations of some features described in Section~\ref{sec:observable_features} and its subsections, and measured by 2D fitting (Section~\ref{sec:2d_fitting}).

In NGC~4535, there two major spiral arms which pass through the CR, marked red and green on the image. First of all, one can see that the general view of spiral structure is different inside and outside the CR: in the inner part of a galaxy, two major arms dominate the spiral structure, whereas outer spiral structure consists of shorter spiral arms and spurs. As we mentioned in Section~\ref{sec:bendings}, any discontinuity in spiral pattern can, in principle, align with the CR location. Also, the OLR is located at 120--140 arcsec which is roughly consistent with the truncation of inner spiral structure, as expected in~\cite{Elmegreen1992}.

Next, for the ``green'' spiral arm, we observe that skewness turns zero very close to the CR, at 69 arcsec; for the ``red'' arm, however, skewness does not reach zero anywhere along it. On the other hand, this spiral arm exhibits a bending, which is, again, located near the CR, at 76 arcsec.

Both major spiral arms also have two brightness dips each in their profiles. Curiously, in both arms each pair of dips is located at similar radii, but neither is fully consistent with CR: the inner pair is located near at 53 and at 57 arcsec, and the outer is at 87 and 89 arcsec. For UV spiral arms, dips are also present at similar radii.

Next, we focus on offsets between spiral arms in IR and UV. Using straightened spiral arms, we measure offsets using slice-by-slice method. For each azimuthal slice in straightened spiral arms, we fit a Gaussians and measure their azimuthal offset. As each slice corresponds to a given radius, one obtains the offset dependence on radius. In the same time, the rotation curve $v(r)$ (for example, one from~\cite{Ponomareva2016}), pattern speed $\Omega_p$ and the timescale of star formation $dt$ together define the theoretical dependence of the offset on radius. However, the offset dependence on $r$ is non-monotonous and far from any possible theoretical curve. The problem is that UV images show numerous clumps, adding noise to the offset image, and thus any quantitative conclusions would be highly unreliable. Qualitatively, we only see that offset is positive at small $r$ and negative at large $r$, and the radius where sign changes is expected to be the location of the CR.

\begin{figure}[H]
\centering
\includegraphics[width=0.99\textwidth]{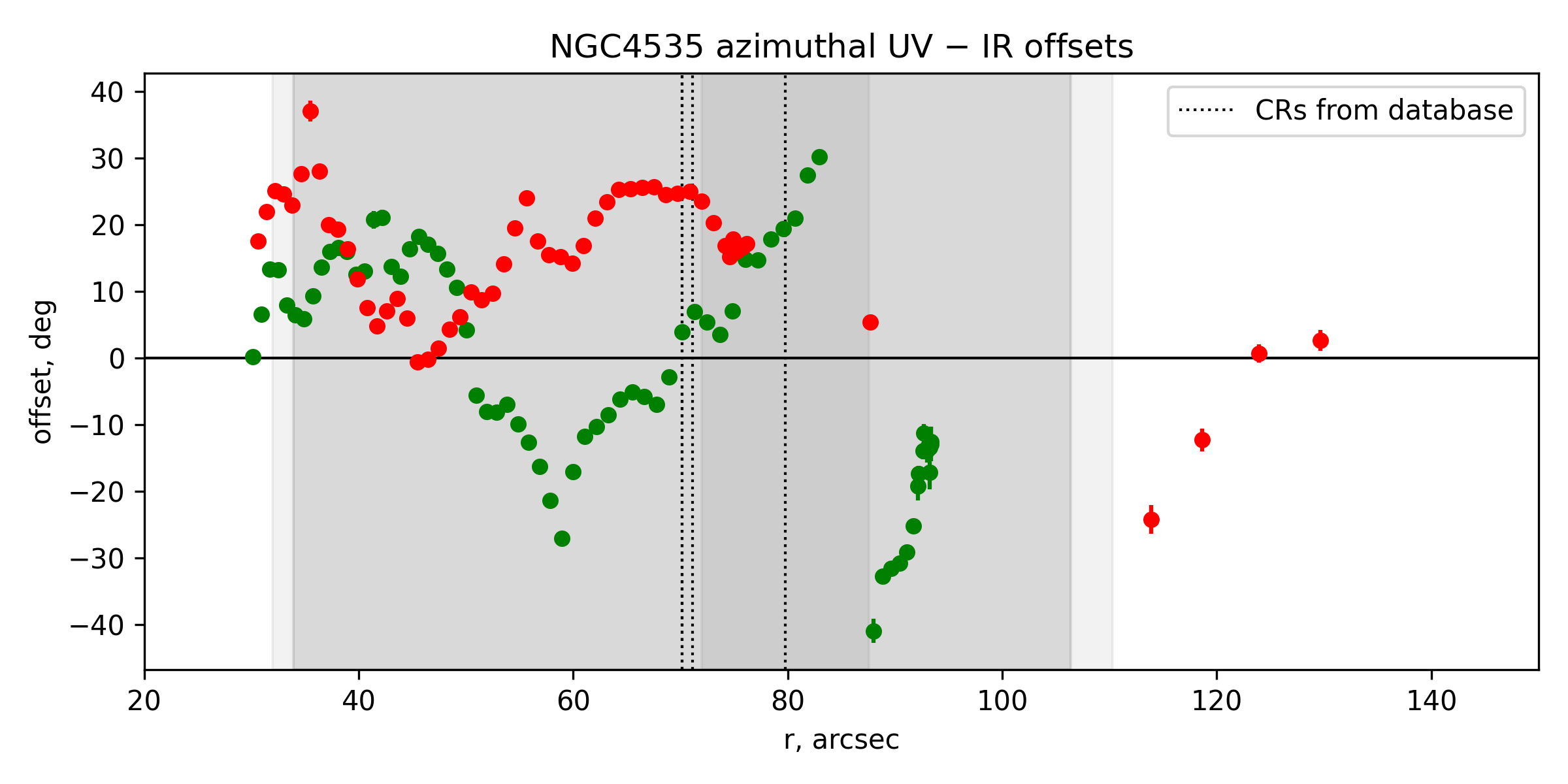}
\caption{This diagram shows the dependence of the UV -- IR azimuthal offset versus radius for two major spiral arms. The locations of measured CRs from database are shown as vertical dotted lines, with uncertainties being shown as grey shaded areas.\label{fig:NGC4535_offsets}}
\end{figure}

However, the observed curve is rather noisy and it deviates significantly from the theoretical prediction. The main issue when analysing spirals in UV images is that clumps dominate the UV radiation, and their random offsets can distort the results. 

Although possible CR indicators are concentrated near the measured CR value, in some cases there are clear inconsistencies, which may be caused by numerous factors. They include image noise or non-smoothness, possible reasons other than the density wave for observed features to appear, or the imperfection of a simple model with a single density wave in the disc. We note that neither of brightness dips features, mentioned in~\cite{Pan2023}, does not align with CR, but there are two pairs of dips just inside and just outside the CR, observed in UV as well as in IR. However, they possibly may be consistent with the location of 4:1 (ultraharmonic) resonance.

In other galaxies, the features in spiral arms discussed before (brightness dips, bendings or radii of zero skewness) are usually less consistent with each other and with any measurements of CRs which had to be expected, as different estimates of the CR may indicate the absence of a single density wave in the disc. Thus, one should not rely on individual CR measurements but employ as many different methods as possible. Fortunately, even purely photometric methods allow one to capture multiple CR indicators. Finally, we note that theoretical modelling indicate that the presence of disc break~\cite{Pohlen2008} can lead to emergence of a pair of spiral modes, not a single one~\cite{Fiteni2024}. Some galaxies in our sample exhibit disc breaks, but not NGC~4535 (see Section~\ref{sec:exp_scales}).

\subsection{Exponential scales of spiral arms}
\label{sec:exp_scales}
\begin{figure}[H]
\centering
\includegraphics[width=0.99\textwidth]{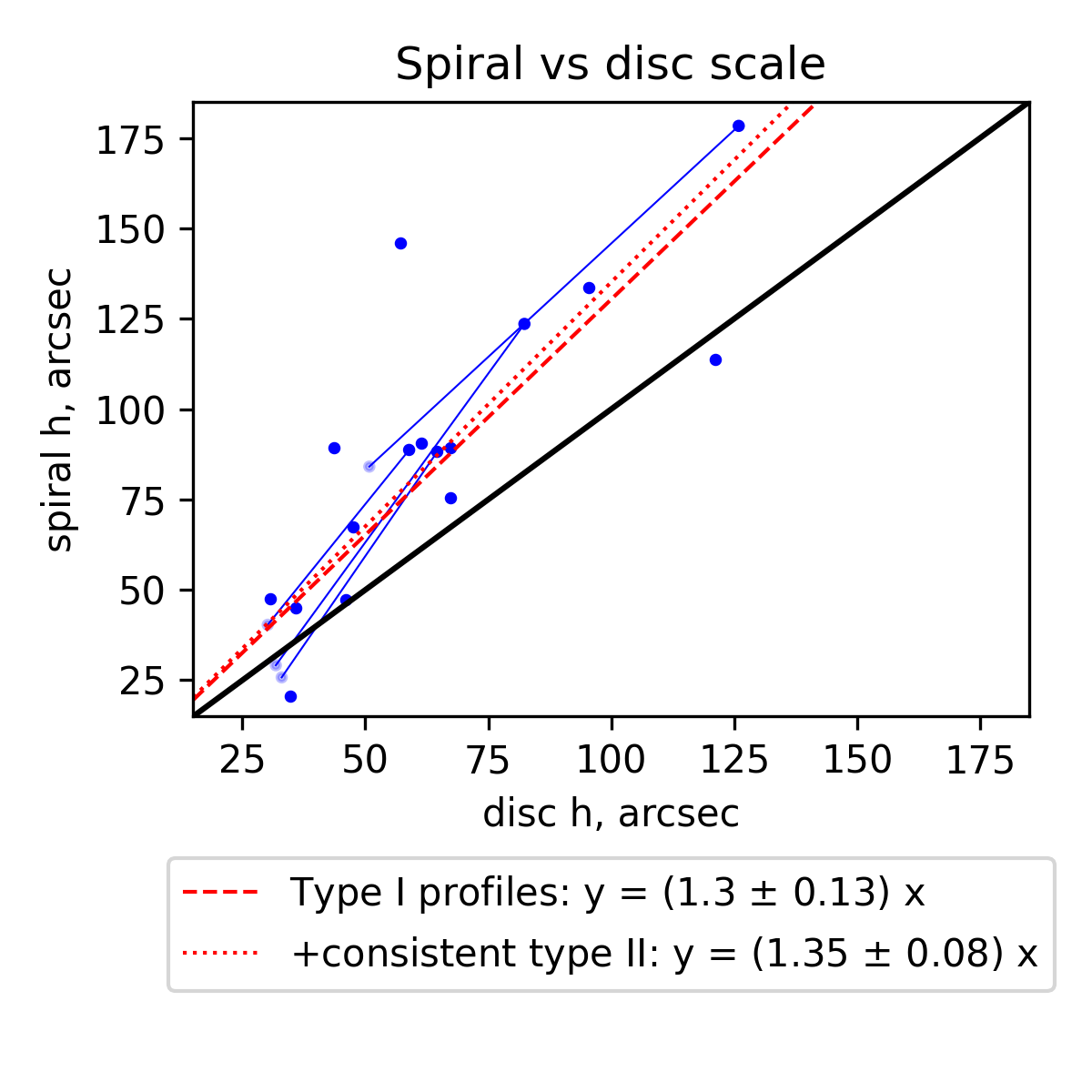}
\caption{A comparison of exponential scales of discs (x-axis) and spiral structures (y-axis) in galaxies. Separate dots represent galaxies which Type I profiles~\cite{Pohlen2008}; pairs of dots connected with lines represent galaxies with Type II profiles, where solid dots represent the inner exponential scale of the disc, and transparent dots represent the outer. Pairs of dots are shown only for galaxies which spirals profile is consistent with Type II disc profile (see text). Dashed and dotted lines represent linear fits of data, solid line shows 1:1 ratio.\label{fig:spiral_exlscales}}
\end{figure}

Despite spiral arms profiles being far from exponential (Section~\ref{sec:I_parallel}), one still can measure their exponential scale and compare spirals with discs by this parameter. We collect all measurements of $I(r)$ for all spirals in a galaxy and fit them with exponential function. For discs, exponential scales were derived from decomposition (Section~\ref{sec:extraction}). Broadly speaking, we found that the exponential scale of the entire spiral structure is of the same order as exponential scale of the disc, or slightly larger (see Figure~\ref{fig:spiral_exlscales}).

More specifically, 7 out of 19 galaxies have breaks in their radial brightness profiles of discs~\cite{Pohlen2008} (all breaks are down-bending, i.e. these galaxies are of type II profiles). Therefore, their discs exhibit two different exponential scales in their inner and outer parts. If we simply exclude them, and only consider galaxies with type I profiles, we observe that spiral arms have larger exponential scales than discs, by 30\% on average.

Next, we consider galaxies with type II profiles. In all cases spiral structure spans simultaneously in the inner and outer parts of the disc. We fit spiral profiles with broken exponential function, similar to the disc; we find that in 4 out of 7 cases, the fitted spiral profile is roughly consistent with the disc profile, i.e. the located break radius of spiral profile is within 20\% of the break radius of disc profile, and the outer scale is larger than the inner. If we add scales for these 4 galaxies to our comparison, we observe that the trend for them generally agrees with the trend for type I profiles: the exponential scales for spiral arms become 35\% larger than for discs. We note that results concerning exponential scales of spiral arms remain the same qualitatively, if we extract spiral arms subtracting the model disc from decomposition with mask, as described above. In this case, spiral arms exponential scales are 33\% larger than disc exponential scales, given that only type I profiles are considered, or 38\% if consistent type II profiles (which stay for the same 4 galaxies) are taken into account.

Our observation that the exponential scale of spiral structure is larger then of disc implies that the contribution of spiral structure to the galaxy luminosity increases with radius. Indeed, in~\cite{Kendall2015} the behaviour of $m = 2$ Fourier component of spiral galaxies images was examined, and one can observe its increase towards the periphery (Figure 1 in their work). However, the analysis of spiral arm contribution as a function of radius in~\cite{Chugunov2024} has shown that it usually peaks at less than $3h$, implying that after this radius spiral brightness tends to fall off faster. Spiral arms truncation is even expected due to the termination of star formation at low gas densities~\cite{Kregel2004}. Finally, the results of~\cite{Gao2017} can be expressed as the exponential scale of the underlying part of disc is larger than of the disc and spiral arms combined, also seemingly inconsistent with our result. However, our result can be explained by the fact that our analysis does not consider truncation, as we fitted only these points where spiral arm flux was measured. Therefore, the combination of all these results can be interpreted as that spiral structure have more flat radial brightness gradient than disc at some range of radii, but eventually truncates, which also highlights that the exponential function is a poor approximation of the overall spiral arm radial brightness profile.

\section{Discussion and conclusions}

Concerning the shape of spiral arms, we confirm that the simple logarithmic spiral shape is not suitable to fit most of spiral arms. Instead, we recommend using polynomial-logarithmic spiral with $N = 2, 3$ or 4, depending on the spiral arm length. Bendings can be added, if needed; they present in 35\% spiral arms, albeit sometimes they are not clearly visible in images. We confirm that this model is enough to fit the majority of spiral arms and, in the same time, it does not introduce overfitting. Our measurements of the average pitch angle are consistent with literature. We examined how properties of spiral arms depend on spiral structure type (grand-design or multi-armed), but did not found any difference which is statistically significant.

Theoretical works usually consider logarithmic spiral arms with the constant pitch angle and do not pay attention to their variations. Moreover, we confirm that spiral arm pitch angles tend to not only vary, but also decrease to the periphery of the galaxy. It implies that representation of spiral arms as logarithmic spiral not only overly simplifies their shape, but also introduces a bias by not accounting this trend.

In the first time, using consistent direct measurements of exact arm form and position, we found that spiral arm radial brightness profile is often far from exponential, exhibiting non-monotonous exponential scale in different parts. In 24\% of arms, we observe pronounced dips in brightness. 
In the same time, the exponential scale of the spiral structure in general is usually slightly larger than the exponential scale of the disc. 

The width of spiral arms usually varies with radius. We found that the linear function of radius describes this variation well, and the constant term (zero-point) is less important than linear. In other words, a width strictly proportional to radius is a better simplification for a model, than a constant width. We emphasise that the absolute width of spiral arm depends significantly on the convention, how spiral arms and disc should be separated.

Combining all above, we propose an analytical function capable to describe the surface brightness profile of spiral arms. We validate this function and confirm our results with 2D fitting of straightened spiral arms, checking the importance of individual parameters. We identify the least important parameters which can be omitted when dealing with spurs, or when highly precise fits are not needed.

We also employ our methods to examine a connection between photometric features, such as brightness dips and locations of zero skewness with corotation radii, which are one of the most important indicators of the nature of spiral arms. In particular, we check our results for NGC~4535, for which we suspect the presence of a density wave, and our findings align with this interpretation.

We observe that ultraviolet images are difficult to fit with smooth functions due to the chaotic and clumpy nature of star formation. As the radiation in most of optical and near-IR bands is emitted by the combination of old and young stellar population, this problem of clumps contamination can possibly arise in different wavelengths. In the recent work~\cite{Kalita2025b}, photometric decomposition of distant galaxies including clumps was performed. Noteworthy, the authors mention that more than 70\% of clumps are associated with spiral features. Considering this, probably the most optimal way to fit galaxies with spiral structure is to combine our approach with one from their work. Modelling spiral arms as a combination of smooth large-scale structure and small clumps can be useful to obtain more precise representation of light distribution in galaxies, and to discern qualitatively different sources of radiation.

More broadly, this work highlights the lack of concordance on the question, how spiral structure should be treated in photometry: should it be interpreted as a feature \textit{above} the disc (and therefore, inter-arm region without spiral features, is indeed a proper disc), or as a some kind of a disturbance inseparable from the disc, manifesting itself as a combination of positive and negative deviations of brightness from the axisymmetric disc. Essentially, this difference in approach results in the different estimates of spiral arm widths (between this work and~\cite{Savchenko2020}), and, in other works, in different contributions of spiral structure into the total luminosity of the galaxy (\cite{Chugunov2024} versus~\cite{Savchenko2020}). Also, these different points of view are manifested in different approaches to decomposition with spiral structure. On the one side, there are some works, including~\cite{Lingard2020} or our works starting from~\cite{Chugunov2024}, where spiral arms are modelled as separate entities; on the other hand, there is a popular approach implemented in GALFIT, where spiral structure can be modelled as a disc, modified with Fourier and bending modes~\cite{Peng2010}. Probably, the answer to this question is connected to the ultimate problem of the nature of spiral arms: are they indeed mostly density waves, or not? If the former is true, then considering spirals as separate features, existing above the disc, does not have much physical meaning. And if not, another question arises: how to separate spirals from the disc correctly?

But, nevertheless, the main result of this work remains: we have constructed and justified a 2D photometric function of the spiral arm. This function can be used for photometric decomposition, which is a powerful tool to retrieve the parameters of spiral arms. The approach of decomposition with spiral arms was already implemented to study the distribution of dust in different components of a galaxy~\cite{Marchuk2025}, to examine the physical nature of spiral arms in different galaxies~\cite{Marchuk2024b, Kostiuk2025}, to find how spiral structure evolves with time~\cite{Chugunov2025}, and so on. In some recent works, for example~\cite{Kalita2025a}, decomposition method, if used, could also be advantageous for the key idea of the study. Now, when a justified photometric model is presented, such decomposition studies can be streamlined and we can expect their results to be more reliable in the future.

\vspace{6pt} 



\authorcontributions{Conceptualization, I.V.C. and A.A.M.; methodology, I.V.C., A.A.M., S.S.S.; software, I.V.C.; validation, I.V.C., A.A.M., S.S.S.; writing---original draft preparation, I.V.C.. All authors have read and agreed to the published version of the manuscript.}

\funding{This research received no external funding.}

\dataavailability{The raw data supporting the conclusions of this article is made available in \url{https://github.com/IVChugunov/Spiral_shapes_and_profiles}.}

\acknowledgments{None.}

\conflictsofinterest{The authors declare no conflicts of interest.}

\abbreviations{Abbreviations}{
The following abbreviations are used in this manuscript:\\
\noindent 
\begin{tabular}{@{}ll}
BIC & Bayesian Information Criterion\\
CR & Corotation radius\\
FWHM & Full width at half-maximum\\
HWHM & Half-width at half-maximum\\
IR & Infrared\\
PSF & Point spread function\\
UV & Ultraviolet
\end{tabular}
}

\begin{adjustwidth}{-\extralength}{0cm}

\reftitle{References}


\bibliography{template}

%


\PublishersNote{}
\end{adjustwidth}
\end{document}